\documentclass[journal,draftclsnofoot,onecolumn,12pt,twoside]{IEEEtran}
\usepackage{cite}

\ifCLASSINFOpdf
  \usepackage[pdftex]{graphicx}
\else
\fi
\ifCLASSOPTIONcompsoc
  \usepackage[caption=false,font=normalsize,labelfont=sf,textfont=sf]{subfig}
\else
  \usepackage[caption=false,font=footnotesize]{subfig}
\fi

\usepackage{stfloats}
\usepackage{epstopdf}
\usepackage{xcolor}
\usepackage{amsmath,amsthm,amsfonts,amssymb}
\usepackage{multirow}
\usepackage{diagbox}
\usepackage{color}
\usepackage{comment}

\begin{document}
%
\title{LTE in Unlicensed Bands is \\neither Friend nor Foe to \mbox{Wi-Fi}}
%
%
%

\author{Ljiljana~Simi\'c,
        Andra~M. Voicu,
        Petri~M\"ah\"onen,
        Marina~Petrova,
        and~J.~Pierre~de~Vries
\thanks{L.~Simi\'c,
        A.~M. Voicu,
        P.~M\"ah\"onen,
        and M.~Petrova are with the Institute for Networked Systems, RWTH Aachen University.}
\thanks{J.~P.~de~Vries is with Silicon Flatirons Centre, University of Colorado.}}
\maketitle

\begin{abstract}
Proponents of deploying LTE in the 5 GHz band for providing additional cellular network capacity
have claimed that LTE would be a better neighbour to \mbox{Wi-Fi} in the unlicensed band, than \mbox{Wi-Fi} is to
itself. On the other side of the debate, the \mbox{Wi-Fi} community has objected that LTE would be highly
detrimental to \mbox{Wi-Fi} network performance. However, there is a lack of transparent and systematic
engineering evidence supporting the contradicting claims of the two camps, which is essential for
ascertaining whether regulatory intervention is in fact required to protect the \mbox{Wi-Fi} incumbent from
the new LTE entrant. To this end, we present a comprehensive coexistence study of \mbox{Wi-Fi} and
\mbox{LTE-in-unlicensed}, surveying a large parameter space of coexistence mechanisms and a range of
representative network densities and deployment scenarios. Our results show that, typically,
harmonious coexistence between \mbox{Wi-Fi} and LTE is ensured by the large number of 5 GHz channels.
For the \mbox{worst-case} scenario of forced \mbox{co-channel} operation, LTE is sometimes a better neighbour to
\mbox{Wi-Fi}---when effective node density is low---but sometimes worse---when density is high.
We find that distributed interference coordination is only necessary to prevent a ``tragedy of the
commons'' in regimes where interference is very likely. We also show that in practice it does not
make a difference to the incumbent what kind of coexistence mechanism is added to
\mbox{LTE-in-unlicensed}, as long as one is in place. We therefore conclude that LTE is neither friend nor
foe to \mbox{Wi-Fi} in the unlicensed bands \emph{in general}. We submit that the systematic engineering analysis
exemplified by our case study is a \mbox{best-practice} approach for supporting \mbox{evidence-based} rulemaking
by the regulator.
\end{abstract}


%
\IEEEpeerreviewmaketitle

\section{Introduction}
%
%
%
%

 

Deploying LTE in the unlicensed 5 GHz band has recently emerged as a way to address growing
data traffic volumes in cellular networks. Proponents have gone so far as to assert that LTE in the
unlicensed bands would not only deliver better performance to its own users than relying on \mbox{Wi-Fi}
offloading, but that it would also improve the performance of existing \mbox{Wi-Fi} networks~\cite{lteuforum2015, qualcomm2015}.
Namely, \mbox{LTE-in-unlicensed} is claimed to be a better unlicensed neighbour to \mbox{Wi-Fi} than \mbox{Wi-Fi} is to
itself. On the other side of what has become a heated debate, the \mbox{Wi-Fi} community has objected that
LTE will be detrimental to \mbox{Wi-Fi} networks~\cite{wifialliance2015}, i.e. that LTE will not be a friend, but rather a foe to
\mbox{Wi-Fi}. (The potential for \mbox{LTE-in-unlicensed} to be used in \mbox{anti-competitive} ways is beyond the scope
of this paper.) Unfortunately, there has been a lack of transparent, reproducible, systematic
engineering analysis submitted as evidence of such claims, by either lobbying camp.

Various mechanisms have been proposed to ensure harmonious coexistence of LTE with \mbox{Wi-Fi}, the
key variants being the LBT (\mbox{listen-before-talk}) based LAA (license assisted access)~\cite{3gpp2015} and the \mbox{duty-cycle}
based \mbox{LTE-U}~\cite{lteuforum2015}. However, the limited industry reports (e.g.~\cite{lteuforum2015}) evaluating the impact of
\mbox{LTE-in-unlicensed} variants on \mbox{Wi-Fi} are very vague on the exact assumptions and scenarios
underlying their results, whereas existing academic research (see e.g. literature review in~\cite{voicu2016})
typically focuses on optimizing specific aspects of coexistence mechanisms. Consequently, the
Federal Communications Commission (FCC) in the US has taken the unusual step---given its
traditionally strong \mbox{technology-neutral} stance---of issuing a Public Notice~\cite{fcc2015} requesting comments
on how \mbox{LTE-in-unlicensed} will coexist with other technologies, including \mbox{Wi-Fi}.

The burning public policy question is whether \emph{regulatory action} is required to protect \mbox{Wi-Fi}, as the
soft incumbent\footnote{We adopt the term \emph{incumbent} to reflect the status of \mbox{Wi-Fi} as the dominant current technology in the unlicensed band, and the resulting expectation of \mbox{Wi-Fi} stakeholders that its operation should not be degraded by an entrant technology beyond what further
\mbox{Wi-Fi} densification would do. However, both LTE entrants and \mbox{Wi-Fi} incumbents have \emph{equal} rights in the unlicensed band.} 
of the unlicensed band, from the new entrant LTE technology. Specifically, key
engineering questions that the spectrum regulator needs answered to make a reasoned policy decision
are: (i)~under what circumstances is \mbox{LTE-in-unlicensed} a friend or a foe to \mbox{Wi-Fi}; (ii)~does the
\mbox{LTE-in-unlicensed} coexistence mechanism, e.g. \mbox{LTE-U} vs. LAA, matter; (iii)~how sensitive are the
conclusions to the network scenarios considered; and (iv)~what is the typical vs. \mbox{worst-case}
coexistence performance?

In this paper, we study coexistence of \mbox{Wi-Fi} with a variety of candidate \mbox{LTE-in-unlicensed} entrants
for different network densities and deployment scenarios, in order to identify whether or where
issues warranting regulatory intervention may arise. Our results show that, for realistic network
densities, \mbox{near-perfect} coexistence between \mbox{Wi-Fi} and \mbox{LTE-in-unlicensed} is ensured simply by virtue
of the large number of 5 GHz channels, so that co-channel operation is easily avoided. For the \mbox{worst-case}
scenario of forced \mbox{co-channel} operation---or equivalently, locally \mbox{higher-than-typical} network
density---LTE is sometimes a better neighbour to \mbox{Wi-Fi} than a \mbox{Wi-Fi} entrant, but sometimes worse.
Beyond this LTE/\mbox{Wi-Fi} case study, we argue that such a systematic engineering analysis, exploring a
large design parameter space~\cite{mahonen2012} for entrant technologies and scenarios, is a \mbox{best-practice} approach \emph{in
general} for supporting \mbox{evidence-based} rulemaking by the regulator.

\section{Methodology \& Scenarios}

In our case study an entrant \mbox{LTE-like} system coexists with incumbent \mbox{Wi-Fi} operation in the 5~GHz
unlicensed band. We assume the incumbent access points\footnote{We use the term ``access point'' to refer both \mbox{Wi-Fi} APs and \mbox{LTE-in-unlicensed} base stations, since both belong to \mbox{small-cell} infrastructure networks with similar functionalities.} 
(APs) and their associated users are
always \textbf{\emph{\mbox{Wi-Fi}}} devices, implementing the PHY layer of IEEE~802.11n and LBT\footnote{Since the CSMA/CA MAC mechanism implemented by IEEE 802.11 \mbox{Wi-Fi} is a version of the more general LBT mechanism, we also refer to it as LBT.}
with random binary exponential back off at the MAC layer~\cite{ieee2012}. In order to broadly yet systematically
evaluate the potential impact on incumbent APs coexisting with likely entrant AP variants in
different deployment scenarios, we explore a large design parameter space for entrant technologies,
and we evaluate the network performance for several distinct deployment scenarios, for different
incumbent and entrant AP densities, as summarized in Table~\ref{table_1}. We perform a coexistence analysis
for each considered case by estimating the \emph{downlink throughput per AP}\footnote{We note that throughput, aside from being the fundamental network performance evaluation metric \emph{in general}, is also considered as the primary performance metric in major LTE/Wi-Fi coexistence studies, e.g.~\cite{lteuforum2015},~\cite{wifialliance2015}.} based on Monte Carlo simulations in MATLAB. For simplicity, we assume each AP (incumbent or entrant) has one associated user and that traffic is downlink and \mbox{full-buffered}. Thus, we report the \mbox{per-AP} throughput
by measuring the throughput of its associated user.\footnote{In the case of multiple users per AP, this \mbox{per-AP} downlink throughput would be split among the users.}

We argue that the systematic engineering analysis exemplified by our case study is \mbox{best-practice} for
facilitating \mbox{evidence-based} discussion among different \mbox{stake-holders} and supporting rulemaking by
the regulator. The key aspect of the methodology we advocate is to describe the proposed technologies, scenarios, performance metrics, and evaluation methods such that the results are
readily reproducible by third parties. Moreover, the analysis ought to be conducted using the best
publicly available knowledge and consider a broad parameter space at an abstraction level that
provides sufficiently accurate yet generalizable conclusions.

We note that for evaluating the regulatory implications of potential coexistence issues during the pre-standardization stage of new entrant technologies like \mbox{LTE-in-unlicensed}, it is also crucial to select a \emph{scalable} analysis model which can produce generalizable results that characterize the proposed technology at the network level. At such a stage any existing entrant devices implement \mbox{pre-standard} or proprietary algorithms and are typically not available for commercial use or testing. Therefore, no decision of general interest and relevance can be made solely based on illustrative \mbox{proof-of-concept} measurements, e.g.~\cite{qualcomm2015}, using such devices. It follows that a \mbox{measurement-based} methodology \emph{alone} cannot answer the public policy question of whether regulatory intervention is required to ensure harmonious coexistence between LTE and \mbox{Wi-Fi} in the unlicensed bands, given the \mbox{network-wide} effects that should be taken into account for various candidate \mbox{LTE-in-unlicensed variants}, representative scenarios, and network densities. We thus argue that to support \mbox{evidence-based} rulemaking, it is imperative to use scalable models that survey a large parameter space and enable a comparative and transparent engineering analysis.

\subsection{Entrant Technologies \& Coexistence Mechanisms}
\label{ent_tech}

We model all entrant technologies by varying the relevant key parameters at the PHY and MAC
layers, as specified in Table~\ref{table_1}. We consider six variants for the entrant APs and users. The first three
entrant variants represent the key flavours of LTE-in-unlicensed that have emerged in various
industry and research discussions on LAA~\cite{3gpp2015} and \mbox{LTE-U}~\cite{lteuforum2015}: \textbf{(i)~\emph{LAA}} with an LTE PHY layer and
LBT at the MAC layer; \textbf{(ii)~\emph{\mbox{LTE-U} fixed 50\% duty cycle}} with an LTE PHY layer and a fixed 50\%
duty cycle MAC; and \textbf{(iii)~\emph{\mbox{LTE-U} adaptive duty cycle}} with an LTE PHY layer and an adaptive duty
cycle MAC, based on the number of other entrant and incumbent APs detected within the carrier
sense (CS) range.

Additionally, we consider the entrant variant \textbf{(iv)~\emph{\mbox{LTE-U} ideal}} to represent the upper bound of
coordination achievable among \mbox{LTE-in-unlicensed} entrants, equivalent to a perfect TDMA schedule
among APs within CS range of each other.\footnote{\emph{\mbox{LTE-U} ideal} is thus a variant of \emph{\mbox{LTE-U} adaptive duty cycle} where transmissions from entrant APs within each other's CS range never occur simultaneously (\emph{cf.} possibility of \mbox{entrant-entrant} interference given uncoordinated transmissions in \emph{\mbox{LTE-U} fixed} and \emph{\mbox{LTE-U} adaptive} variants).} 
We also consider \textbf{(v)~\emph{LTE}} as the most basic LTE entrant
variant, which transmits continuously and implements no mechanism to facilitate coexistence with
the \mbox{Wi-Fi} incumbent. Finally, to help us answer the question of whether these \mbox{LTE-like} entrant
variants are better neighbours to \mbox{Wi-Fi} than \mbox{Wi-Fi} is to itself, as has been suggested by proponents of
\mbox{LTE-U}~\cite{qualcomm2015}, we consider the incumbent \textbf{(vi)~\emph{\mbox{Wi-Fi}}} as the reference baseline entrant.

We focus on the combination of \emph{channel allocation} and \emph{MAC} protocol that chiefly\footnote{Any other parameters, e.g. transmit power, that influence the extent of mutual interference could arguably also be considered part of a ``coexistence mechanism'', but their primary function is not to facilitate coexistence.} 
constitutes a \emph{coexistence mechanism} in the case of multiple wireless systems (whether implementing the same
technology or not) sharing a common radio spectrum band. To this end, we also consider different
basic channel selection mechanisms that the incumbent and entrant APs use to select one of the
available 20 MHz operating channels in the 5 GHz band. We assume that the incumbent \emph{\mbox{Wi-Fi}} APs
randomly select an operating channel, whereas the entrant APs use either \textbf{\emph{random}} channel selection,
or \textbf{\emph{sense}} channel selection where the entrant AP selects (randomly) a channel unoccupied by an
incumbent AP.

The total number of available 20 MHz channels in the 5 GHz band in Europe is nineteen for indoor
or eleven for outdoor operation~\cite{ieee2012}. Given this large number of available channels, our considered
scenarios and typical network densities (see Table~\ref{table_1}) result in only one or two APs per channel on
average. We thus expect that \emph{sense} channel selection will not have a large impact on \mbox{incumbent-entrant}
coexistence compared to \emph{random} selection. Therefore, we also consider \textbf{\emph{forced \mbox{co-channel}}}
channel allocation, where all incumbent and entrant APs in the scenario occupy a single 20 MHz
operating channel. The {\emph{forced \mbox{co-channel}} case represents a \mbox{worst-case} interference scenario,
modelling important scenarios with a higher than usual risk of \mbox{inter-system} interference, such as:
locally \mbox{higher-than-typical} network density; restricted availability of channels (e.g. only four
\mbox{non-DFS} channels in Europe); or implementing channel aggregation to operate over fewer
\mbox{higher-bandwidth} channels, as in IEEE 802.11ac. We argue that although such \mbox{worst-case} scenarios
may occur infrequently, they are crucial since they best reflect the situations when coexistence
problems, and possible calls for regulatory intervention, would arise.

\subsection{Deployment Scenarios}

We study entrant and incumbent network coexistence in three major deployment scenarios, as
illustrated in Fig.~\ref{fig_1} and summarized in Table~\ref{table_1}. In the \textbf{\emph{indoor/indoor}} scenario, both the incumbent
and entrant networks are indoors; in the \textbf{\emph{indoor/outdoor}} scenario, the incumbent APs and users are
indoors, whereas the entrants are outdoors; in the \textbf{\emph{outdoor/outdoor}} scenario, both the incumbent and
entrant networks are outdoors.

For all outdoor deployments, we generate the network topology by randomly allocating each
outdoor AP to one of 20 real cellular pico base station locations from central London~\cite{mls2015}. The
associated outdoor users are randomly located within the AP's coverage area and at a maximum
distance of 50 m. For indoor network deployments, we assume the 3GPP dual stripe model~\cite{alcatel2009}.

In the \emph{indoor/indoor} scenario, we assume all incumbent and entrant APs coexist in one \mbox{single-floor}
building with twenty apartments, as shown in Fig.~\ref{fig_1}\subref{fig_1a}. We consider one AP or fewer per apartment,
so that the corresponding indoor network densities of 600--6000 APs/km\textsuperscript{2} are consistent with the
recent \mbox{Wi-Fi} measurements in~\cite{achtzehn2013}. In the \emph{indoor/outdoor} scenario, as per the 3GPP dual stripe
model, \mbox{multi-floor} buildings are randomly overlayed\footnote{We note that although our modelling of the urban layout is thus a simplification of a real (typically more regular) city layout, this is not expected to significantly affect our results given the \mbox{ITU-R} LOS/NLOS outdoor propagation model (\emph{cf.} Section~\ref{sim_model}).} 
on the study area containing the outdoor AP locations, as shown in Fig.~\ref{fig_1}\subref{fig_1b}. The incumbent APs and users are randomly located inside the
apartments with a density of 500--5000 APs/km\textsuperscript{2}, whereas we consider one to 20 outdoor entrant APs
(corresponding to an outdoor network density of 7--150 APs/km\textsuperscript{2}). In the \emph{outdoor/outdoor} scenario,
we consider up to 20 outdoor entrant and incumbent APs in total (corresponding to an outdoor
network density of 14--150 APs/km\textsuperscript{2}), as shown in Fig.~\ref{fig_1}\subref{fig_1c}.

\clearpage

\begin{table}[t]
\caption{Explored parameter space---entrant variants and scenarios}
\label{table_1}
\centering
\begin{tabular}{|p{1.5cm}|p{1.5cm}|c|c|c|}
\hline
	\multicolumn{2}{|c|}{\diagbox[width=4cm, height=2cm]{\textbf{PARAMETER}}{\textbf{SCENARIO}}} 
	& \begin{tabular}{c} \emph{\textbf{Indoor/indoor}}\\ \textbf{(indoor incumbent,} \\ \textbf{indoor entrant)}\end{tabular}
	& \begin{tabular}{c} \emph{\textbf{Indoor/outdoor}}\\ \textbf{(indoor incumbent,} \\ \textbf{outdoor entrant)}\end{tabular}
	& \begin{tabular}{c} \emph{\textbf{Outdoor/outdoor}}\\ \textbf{(outdoor incumbent,} \\ \textbf{outdoor entrant)}\end{tabular} \\
\hline
	\multicolumn{2}{|c|}{\textbf{Network size}}
	& \begin{tabular}{p{1.2cm}p{1.7cm}} \textbf{incumbent}:&1 \emph{or} 10~APs\\ \textbf{entrant}:&1--10~APs\end{tabular}	
	& \begin{tabular}{p{1.2cm}p{2.8cm}} \textbf{incumbent}:&500 \emph{or} 5000~APs/km\textsuperscript{2}\\ \textbf{entrant}:&1--20~APs\end{tabular}
	& \begin{tabular}{p{1.2cm}p{1.7cm}} \textbf{incumbent}:&1 \emph{or} 10~APs\\ \textbf{entrant}:&1--10~APs\end{tabular}\\
\hline
	\multicolumn{2}{|c|}{\textbf{AP transmit power}}	
    & 23~dBm
    & \begin{tabular}{p{1.2cm}p{2.8cm}} \textbf{incumbent}:&23~dBm\\ \textbf{entrant}:&23 \emph{or} 30~dBm\end{tabular}	
    & 23 \emph{or} 30~dBm\\
\hline
	\multicolumn{2}{|c|}{\begin{tabular}{c} \textbf{Maximum number of}\\ \textbf{available channels} \\\textbf{(Europe)} \end{tabular}}
	& 19
	& \begin{tabular}{p{1.2cm}p{2.8cm}} \textbf{incumbent}:&19\\ \textbf{entrant}:&11\end{tabular}	
	& 11\\
\hline
	\multirow{10}{*}{\begin{minipage}{2cm}\emph{\textbf{Coexistence}}\\ \emph{\textbf{mechanism}}\end{minipage}} 
	& \begin{minipage}{2cm}\textbf{Channel \\selection}\end{minipage}
	& \multicolumn{3}{l|}{\begin{tabular}{l} \textbf{incumbent}: random \emph{or} forced co-channel\\
	                     \textbf{entrant}: random \emph{or} sense (select channel with fewest incumbent APs) \emph{or} forced co-channel
	                      \end{tabular}}\\
\cline{2-5}                      
	& \textbf{MAC} & \multicolumn{3}{l|}{\begin{tabular}{p{0.1cm}rl} 
	                                     \textbf{incumbent}: & \emph{\textbf{Wi-Fi}}: & LBT, CS threshold of (-62)-82~dBm for (non-)Wi-Fi devices\\
	                                     \textbf{entrant}: & & \\
	                                     & \emph{\textbf{LAA}}: & LBT, CS threshold of -62~dBm\\
	                                     & \emph{\textbf{LTE-U fixed 50\% duty cycle}}: &ON/OFF with 50\% duty cycle\\
	                                     & \emph{\textbf{LTE-U adaptive duty cycle}}: &ON/OFF with adaptive duty cycle based on number of entrant\\
	                                     & & \& incumbent APs within CS range (CS threshold = -62~dBm)\\
	                                     & \emph{\textbf{LTE-U ideal}}: & ideal TDMA\\
	                                     & \emph{\textbf{LTE}}: & always ON (continuous transmission)\\
	                                     & \emph{\textbf{Wi-Fi}}: &LBT, CS threshold of (-62)-82~dBm for (non-)Wi-Fi devices
	                                    \end{tabular}}\\
\hline
	\multicolumn{2}{|c|}{\textbf{PHY}} &  \multicolumn{3}{l|}{\begin{tabular}{p{0.1cm}rl} 
	                                     \textbf{incumbent}: & \emph{\textbf{Wi-Fi}}: & IEEE 802.11n spectral efficiency $\rho_{WiFi}$, noise figure NF=15~dB\\
	                                     \textbf{entrant}: & & \\
	                                     & \emph{\textbf{LAA}}: & LTE spectral efficiency $\rho_{LTE}$, NF=9~dB\\
	                                     & \emph{\textbf{LTE-U fixed 50\% duty cycle}}: & LTE spectral efficiency $\rho_{LTE}$, NF=9~dB\\
	                                     & \emph{\textbf{LTE-U adaptive duty cycle}}: & LTE spectral efficiency $\rho_{LTE}$, NF=9~dB\\
	                                     & \emph{\textbf{LTE-U ideal}}: & LTE spectral efficiency $\rho_{LTE}$, NF=9~dB\\
	                                     & \emph{\textbf{LTE}}: & LTE spectral efficiency $\rho_{LTE}$, NF=9~dB\\
	                                     & \emph{\textbf{Wi-Fi}}: & IEEE 802.11n spectral efficiency $\rho_{WiFi}$, NF=15~dB 
	                                    \end{tabular}}\\   
\hline      
    \multicolumn{2}{|c|}{\begin{tabular}{c} \textbf{LBT parameters}\\ \textbf{\& assumptions} \end{tabular}} 
           & \multicolumn{3}{|l|}{\begin{tabular}{l} binary exponential random backoff with $CW_{min}$=15, $CW_{max}$=1023,\\
                                    time slot duration $\sigma$=9~$\mu$s, SIFS=16~$\mu$s, DIFS=SIFS+2$\sigma$=34~$\mu$s (\emph{cf}. IEEE 802.11) 		   											\end{tabular}}  \\
\hline
    \multicolumn{2}{|c|}{\begin{tabular}{c} \textbf{LBT frame}\\ \textbf{duration $T_f$} \end{tabular}}                                                           
    & \multicolumn{3}{|l|}{\begin{tabular}{rl} \emph{\textbf{Wi-Fi}}: & $T_f=fn(rate,~MSDU,~PHY_{header},~MAC_{header})$, \\
                                                                      & $MSDU$=1500~Bytes, $PHY_{header}$=40~$\mu$s, $MAC_{header}$=320~bits (\emph{cf}. IEEE 802.11)\\
                                                \emph{\textbf{LAA}}:  & $T_f$=1~ms (i.e. duration of LTE subframe) \end{tabular}}\\
\hline
     \multicolumn{2}{|c|}{\textbf{Duty cycle ON-time}} 
     &  \multicolumn{3}{|l|}{\emph{\textbf{LTE-U}} \textbf{variants}: 100~ms (i.e. maximum ON-time specified in~\cite{qualcomm2015})}\\                                 \hline
     \multicolumn{2}{|c|}{\textbf{User distribution}}
     &  \multicolumn{3}{|l|}{1 user per AP}\\
\hline
     \multicolumn{2}{|c|}{\textbf{Traffic model}}
     &  \multicolumn{3}{|l|}{downlink full-buffered}\\  
\hline
     \multicolumn{2}{|c|}{\textbf{Channel bandwidth}}
     &  \multicolumn{3}{|l|}{20~MHz}\\ 
\hline
     \multicolumn{2}{|c|}{\textbf{Frequency band}}
     &  \multicolumn{3}{|l|}{5~GHz (5150--5350 and 5470--5725 MHz)}\\ 
\hline      
\end{tabular}
\end{table}

\clearpage

\begin{figure}[t!]
\centering
\captionsetup[subfigure]{width=\linewidth}
\subfloat[\emph{Indoor/indoor} scenario: the incumbent and entrant networks are located inside a single-floor building with 20 apartments
(each of 10~m~$\times$~10~m~$\times$~3~m). Each AP and its associated user are randomly placed in a single apartment.]
          {\includegraphics[scale=0.46]{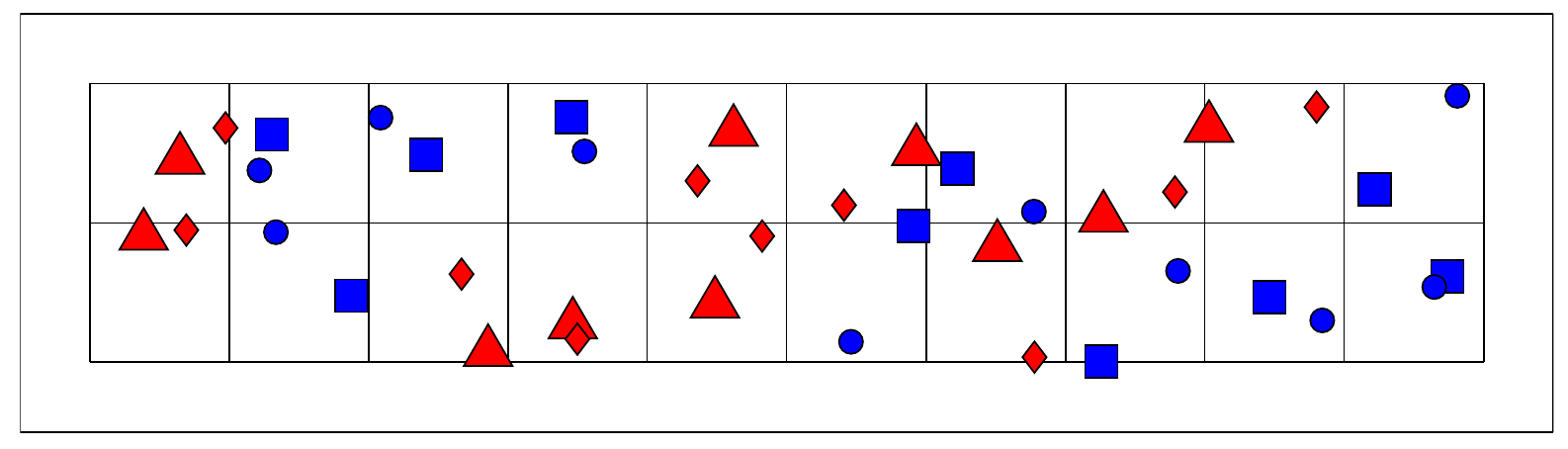} \label{fig_1a}}
\\
\subfloat[\emph{Indoor/outdoor} scenario: the incumbent APs and users are located indoors and the entrant APs and users are located
outdoors. The outdoor entrant users are located in the coverage area of and at a maximum distance of 50~m from the AP
that they are associated with. The length of the buildings is randomly selected between 3--10 apartments and the height is
randomly selected between 3--5 floors. The size of the total study area is 346~m~$\times$~389~m, corresponding to the area in
London where the real locations of the outdoor entrant APs were observed.]   
	  	   {\includegraphics[scale=0.5]{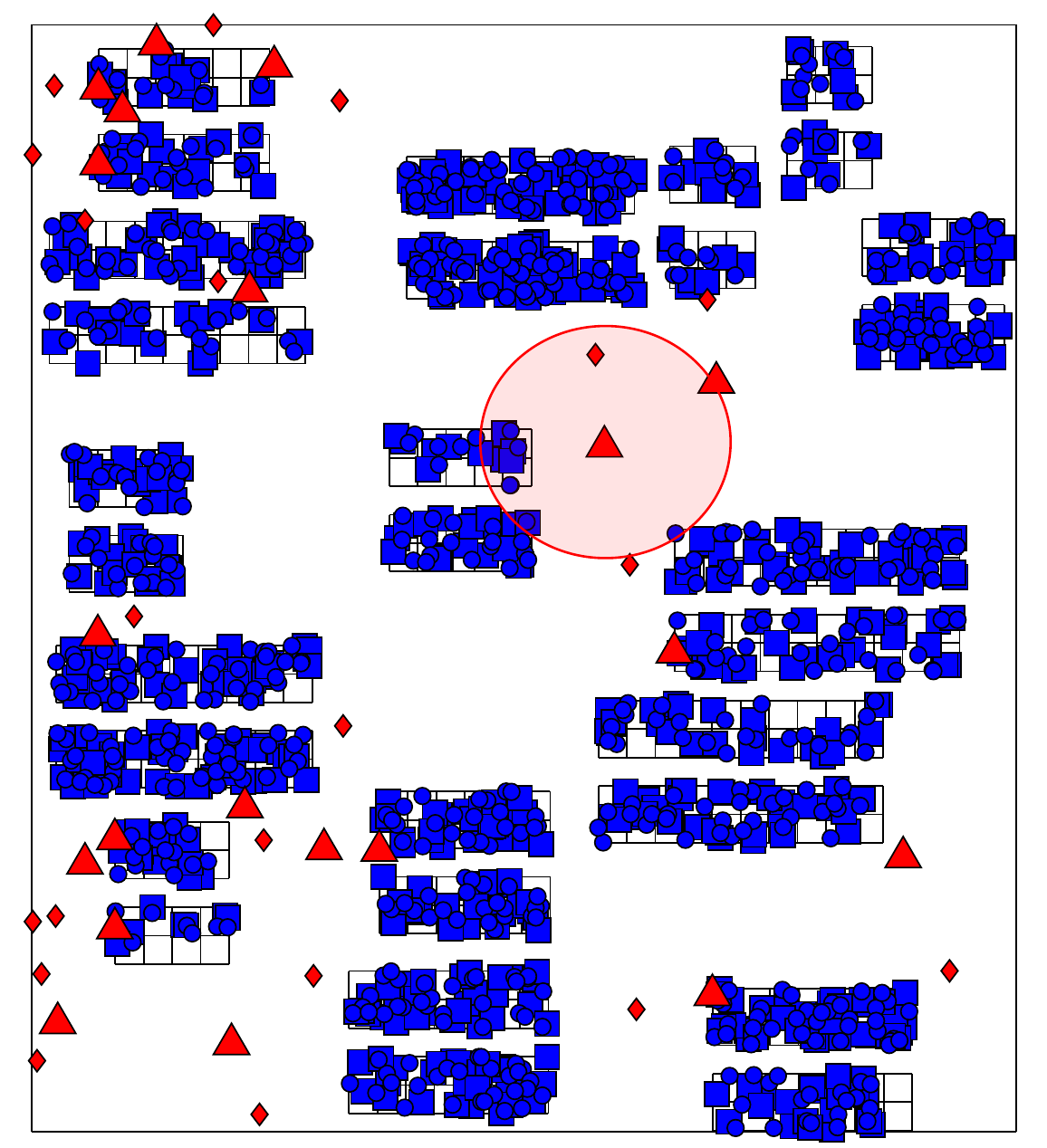} \label{fig_1b}}
\\
\subfloat[\emph{Outdoor/outdoor} scenario: the incumbent and entrant APs are randomly allocated one real outdoor location. The outdoor
users are located in the coverage area of and at a maximum distance of 50~m from the AP that they are associated with.]
          {\includegraphics[scale=0.5]{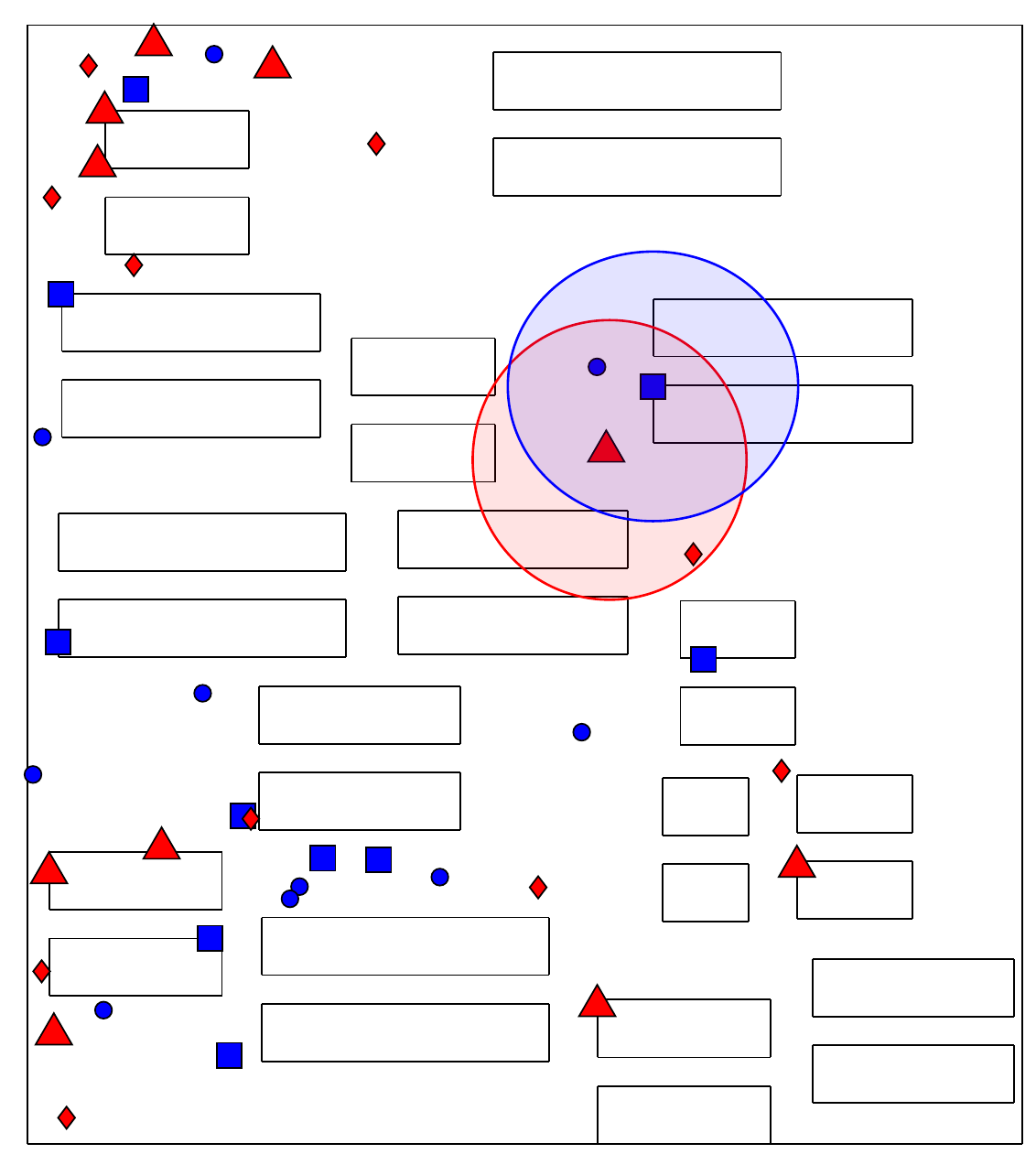} \label{fig_1c}}
\caption{Network layout based on the 3GPP dual stripe model for indoor deployments and real outdoor
picocell locations for outdoor deployments, for the (a)~\emph{indoor/indoor}, (b)~\emph{indoor/outdoor}, and
(c)~\emph{outdoor/outdoor} scenarios, showing example locations of incumbent APs ($\color{blue} \blacksquare$), incumbent
users ($\color{blue} \bullet$), entrant APs ($\color{red} \blacktriangle$), and entrant users ($\color{red} \blacklozenge$).}
\label{fig_1}
\end{figure}

\clearpage

\section{Simulation \& Throughput Model}
\label{sim_model}

We estimate the performance of the coexisting incumbent and entrant technologies using Monte
Carlo simulations in MATLAB, for 3,000 network realizations for the \emph{indoor/indoor} scenario, and
1,500 network realizations for the \emph{indoor/outdoor} and \emph{outdoor/outdoor} scenarios. We assume
propagation models corresponding to indoor and outdoor deployments in our scenarios. For outdoor
links we assume the \mbox{ITU-R} model for \mbox{line-of-sight} (LOS) propagation within street canyons and the
\mbox{ITU-R} \mbox{non-line-of-sight} (NLOS) model for over \mbox{roof-top} propagation~\cite{itur2013}. For indoor links we assume a
\mbox{multi-wall-and-floor} (MWF) model~\cite{lott2001} with a building entry loss of 19.1~dB for external walls. For
outdoor/indoor links we consider cascaded models of indoor and outdoor propagation models. We
assume \mbox{log-normal} shadowing with a standard deviation of 4~dB for indoor links and 7~dB for all other
links~\cite{3gpp2010}.

In order to help explain our throughput model, which we present in detail in\cite{voicu2016}, 
Fig.~\ref{fig_2} gives example transmission sequence
diagrams when the \emph{\mbox{Wi-Fi}} incumbent coexists with each of our six considered entrant technologies.
For technologies that implement LBT at the MAC layer---i.e. \emph{\mbox{Wi-Fi}} and \emph{LAA}---we estimate the
downlink throughput per AP by taking into account other incumbent and entrant APs within CS
range. We assume LBT prevents all \mbox{co-channel} APs within each other's CS range from transmitting
simultaneously and that an AP implementing LBT is granted the channel only for a fraction of time,
while the rest of the time is used by other APs within CS range. \mbox{Co-channel} incumbent or entrant
APs located outside the CS range of the considered AP interfere with this AP by decreasing the
SINR (signal to interference and noise ratio) of its associated user.

For technologies that implement duty cycling at the MAC layer---i.e. \emph{\mbox{LTE-U}} variants---we
estimate the downlink throughput per AP by assuming a slotted time model where all entrant APs
use the same duty cycle \mbox{ON-time} slot duration of 100~ms~\cite{qualcomm2015} and each entrant AP randomly selects a
slot to transmit on (i.e. transmissions from different \mbox{co-channel} entrant APs may overlap in the same
slot). Consequently, for \emph{\mbox{LTE-U} fixed 50\% duty cycle}, the total duration of a duty cycle period is
always equal to two \mbox{ON-time} slots, whereas for \emph{\mbox{LTE-U} adaptive duty cycle}, an entrant AP calculates
a variable duty cycle period as the number of \mbox{ON-time} slots equal to the number of incumbent and
entrant APs within its CS range. Consequently, the throughput of each \emph{\mbox{LTE-U}} entrant is proportional
to its duty cycle period and the SINR of its associated user is decreased by taking into account the
interference from all other \mbox{co-channel} \emph{\mbox{LTE-U}} entrants.

In general, we estimate the downlink throughput\footnote{Our model focuses on the long-term average throughput to characterize the general coexistence performance, and captures the PHY/MAC behaviour of the incumbent and entrant technologies sharing the unlicensed band at a correspondingly appropriate level of complexity and detail; we thus expect other (e.g. \mbox{packet-level}) simulators that capture a finer time granularity to nonetheless provide comparable long-term throughput estimates.} of an incumbent or entrant AP as
\begin{equation}
\label{eq_1}
R=\rho\{SINR\} \times A \times S \times (1-r_{deg}),
\end{equation}
where $\rho$ is the PHY spectral efficiency depending on the SINR of the AP's associated user, $A$ is the
fraction of airtime the AP transmits for~\cite{nguyen2007, simic2012}, $S$ is the LBT MAC efficiency accounting for sensing time and collisions among LBT frames, and $r_{deg}$ is the
degradation of throughput due to collisions between frames from incumbents and any entrants
implementing \emph{\mbox{LTE-U}} duty cycle. The relevant parameters in~(\ref{eq_1}) are specified per entrant variant in
Fig.~\ref{fig_2}, alongside the illustrative examples of coexistence with the incumbent. We adopt the
following notation: $n^{inc}$ and $n^{ent}$ are the number of \mbox{co-channel} incumbent and entrant APs within CS
range, respectively, $m$ is the number of frames transmitted by all incumbents within CS range in one duty cycle \mbox{ON-time} slot,
 $\rho_{WiFi}$ is the IEEE 802.11n spectral efficiency~\cite{ieee2012} mapping the SINR
at the associated user to the throughput, and $\rho_{LTE}$ is the LTE spectral efficiency~\cite{3gpp2009}. We estimate $S$
based on Bianchi's model~\cite{bianchi2000} which we modify in order to take into account parameter values
specific to IEEE 802.11n in the 5~GHz band, and a combination of APs with variable frame duration
depending on the adaptive rate (assumed for \emph{\mbox{Wi-Fi}}) and fixed frame duration (assumed for \emph{LAA})~\cite{voicu2016}.
We note that in order to estimate $n^{inc}$ and $n^{ent}$ we apply the CS threshold values in Table~\ref{table_1}. We also
note that $r_{deg}$ is relevant only for incumbent APs; i.e. for the entrants we do not consider additional
throughput degradation for duty cycle due to collisions with incumbent frames, since the \emph{\mbox{LTE-U}}
entrants have better spectral efficiency and longer transmission time than the \emph{\mbox{Wi-Fi}} incumbents.

The SINR at the incumbent or entrant AP's associated user is given by
\begin{equation}
SINR=fn\{P_{tx}, I^{inc}, I^{ent}, N\},
\end{equation}
where $P_{tx}$ is the transmit power of the AP (given in Table~\ref{table_1} for incumbents and entrants), $I^{inc}$ is the
sum of the \mbox{co-channel} interference from other incumbent APs outside CS range, $I^{ent}$ is the sum of the
\mbox{co-channel} interference from entrant APs, and $N$ is the noise power (where thermal noise is 174~dBm/Hz and NF is given in Table~\ref{table_1}). 
For \emph{LAA}, \emph{\mbox{Wi-Fi}}, and \emph{\mbox{LTE-U} ideal}, $I^{ent}$ is calculated only based
on the \mbox{co-channel} entrant APs outside the CS range, whereas for \emph{LTE}, \emph{\mbox{LTE-U} fixed 50\% duty cycle},
and \emph{\mbox{LTE-U} adaptive duty cycle}, $I^{ent}$ is calculated based on all \mbox{co-channel} entrant APs and on the
\mbox{long-term} average probability that their transmissions overlap with the considered entrant AP.
Finally, we note that if an \emph{LTE} entrant exists within CS range of an incumbent, the incumbent is
always prevented from transmitting since \emph{LTE} is always on, and its throughput is zero.

\clearpage

\begin{figure}[t!]
\centering
\includegraphics[width=\linewidth]{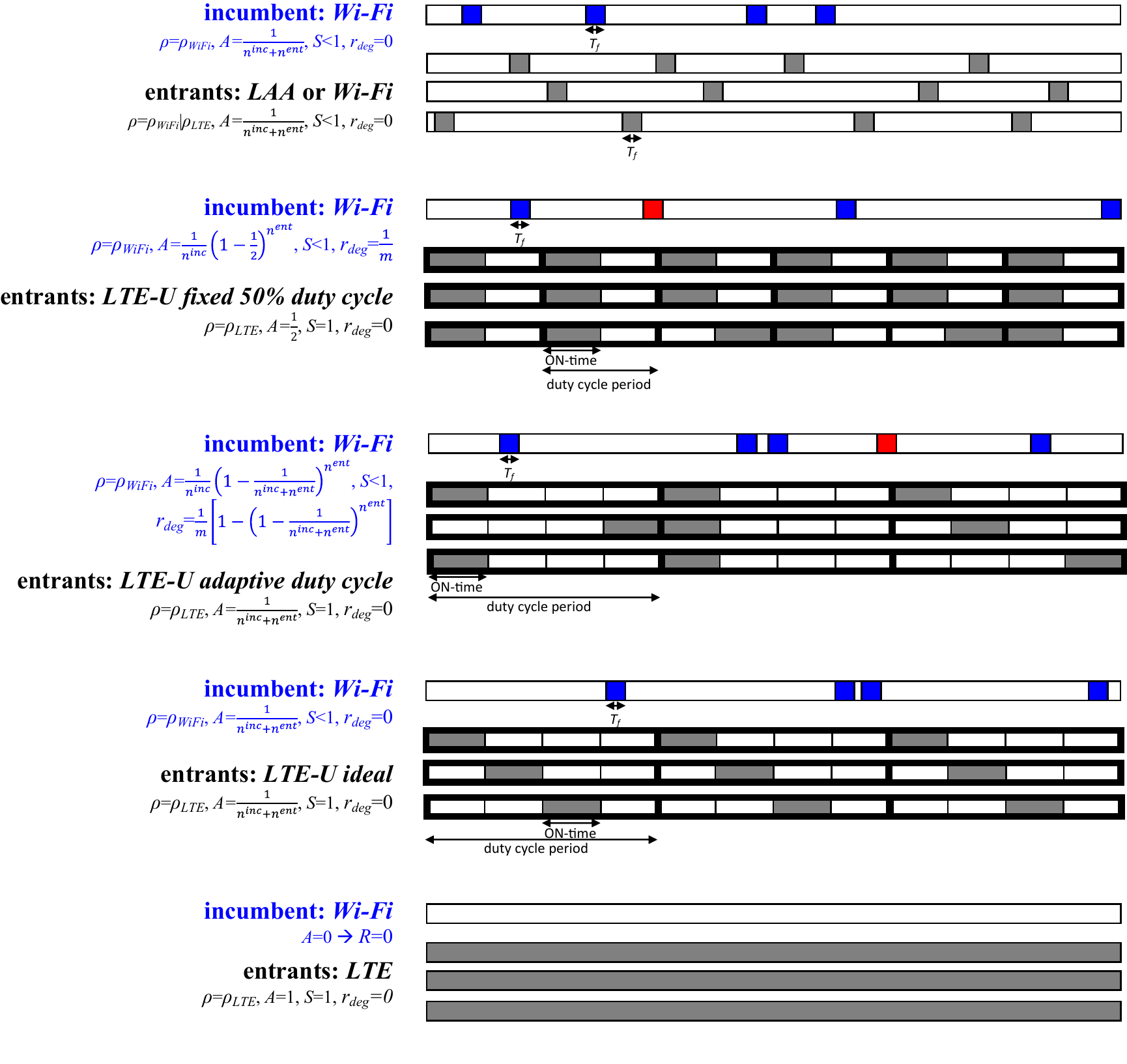}
\caption{Illustration of our throughput model~\cite{voicu2016} for different \mbox{LTE-in-unlicensed} entrant technologies coexisting
with the \emph{\mbox{Wi-Fi}} incumbent, showing the example case of 1 incumbent and 3 entrant APs (all 4 APs are \mbox{co-channel}
and within CS range of each other). AP transmissions are represented as coloured time periods:
\emph{grey} represents entrant AP transmissions, \emph{blue} represents the incumbent AP transmissions that do not
suffer a collision, and \emph{red} represents incumbent frames lost due to collisions with the entrant. During \emph{white}
time periods the AP is not transmitting, in accordance with its MAC coexistence mechanism.}
\label{fig_2}
\end{figure}

\clearpage

\section{Results}
\label{section_results}
\subsection{Typical Impact of Entrant on \mbox{Wi-Fi} Incumbent}

Fig.~\ref{fig_3} shows the \emph{\mbox{Wi-Fi}} incumbent AP throughput when 1 or 10 incumbent APs coexist with an
increasing number of entrant APs in the \emph{indoor/indoor} scenario, for different entrant technologies
using \emph{sense} channel selection. Figs.~\ref{fig_3}\subref{fig_3a} and~\ref{fig_3}\subref{fig_3c} show the median throughput, whereas Figs.~\ref{fig_3}\subref{fig_3b}
and~\ref{fig_3}\subref{fig_3d} show the 10\textsuperscript{th} percentile throughput (over all Monte Carlo network realizations\footnote{Throughout, we present our simulation results in terms of the median and 10\textsuperscript{th} percentile throughput of the ensemble distribution of all Monte Carlo network realizations as specified in Section~\ref{sim_model}, in order to provide both the representative and worst-case throughput experienced by the APs, respectively, in the various considered coexistence scenarios.}). For 1
incumbent AP coexisting with the entrant APs in Figs.~\ref{fig_3}\subref{fig_3a} and~\ref{fig_3}\subref{fig_3b}, a constant throughput of 37~Mbps is achieved regardless of the number of entrant APs or entrant variant. 
For 10 incumbent APs in Figs.~\ref{fig_3}\subref{fig_3c} and~\ref{fig_3}\subref{fig_3d}, a constant throughput of 37~Mbps and 19~Mbps is achieved regardless of the
number of entrant APs or entrant variant, for the median and 10\textsuperscript{th} percentile case, respectively. We
note that we observed the identical trend of constant incumbent throughput also in the
\emph{indoor/outdoor} and \emph{outdoor/outdoor} scenarios when the entrant APs use sense channel selection.\footnote{For the sake of brevity, we focus on explicitly presenting only a representative selection of the results for all studied scenarios in
Table~\ref{table_1}, and discuss our overall simulation results with respect to these.}
Therefore, our results demonstrate that the performance of the \emph{\mbox{Wi-Fi}} incumbent is entirely
independent of the entrant, as long as the entrant implements \emph{sense} channel selection such that it
operates in a channel unoccupied by an incumbent whenever possible. Moreover, our results for the
\emph{indoor/indoor}, \emph{indoor/outdoor}, and \emph{outdoor/outdoor} scenarios using \emph{random} channel selection also
exhibit the same trend of constant median incumbent throughput (and a similar trend for the 10\textsuperscript{th}
percentile throughput).
 
This is an important result: at the realistic network densities considered,
\mbox{near-perfect} coexistence between the \emph{\mbox{Wi-Fi}} incumbent networks and \mbox{LTE-in-unlicensed} entrant
networks is ensured simply by virtue of the large number of available channels in the 5~GHz band, such that \mbox{co-channel} operation is easily avoided.
We note that in practice there may be cases of legacy incumbent \mbox{Wi-Fi} devices that do not perform DFS and thus have a more restricted set of available channels to operate on, i.e. only four non-DFS channels~\cite{fcc2016, etsi2012} (support for DFS is standard in the majority of modern commercial \mbox{Wi-Fi} APs). It remains reasonable to assume that the entrant APs, being at least as sophisticated as modern \mbox{Wi-Fi} devices, will have DFS capabilities and be able to implement \emph{sense}, as required in~\cite{lteuforum2015}. In such a case, the incumbent/entrant coexistence conditions would in fact improve, since perfect isolation between incumbents and entrants would be ensured by the entrant APs always selecting the DFS channels unoccupied by the incumbents. 
For the even less likely case where neither the incumbents, nor the entrants have DFS capabilities (corresponding to rare cases of e.g. faulty DFS implementation by entrant devices), we discuss in Section~\ref{worst-case impact} the worst-case scenario where incumbent and entrant APs locally operate on the same channel, i.e. the \emph{forced co-channel} case.

\subsection{Typical Impact of LTE-in-Unlicensed Entrant Technologies}

Having established that the \mbox{LTE-in-unlicensed} entrant variant makes no difference to the incumbent
throughput in the above scenarios, we now consider what incentives the entrant may have for
adopting different \mbox{LTE-in-unlicensed} technologies. Fig.~\ref{fig_4} presents the throughput of the entrant AP
for the \emph{outdoor/outdoor} scenario with \emph{sense} channel selection. Fig.~\ref{fig_4}\subref{fig_4a} shows that the median
entrant AP throughput is independent of the number of entrant APs, which is consistent with the
avoidance of \mbox{co-channel} operation enabled by the large number of available channels, as observed
for the incumbents in Fig.~\ref{fig_3}. It follows that since there are no other co-channel APs within CS range,
the entrant AP throughput in Fig.~\ref{fig_4}\subref{fig_4a} is simply determined by the \mbox{single-link} PHY and MAC layer
efficiency of the entrant variant. The entrants \emph{\mbox{LTE-U} ideal} and \emph{\mbox{LTE-U} adaptive duty cycle} achieve
the maximum \mbox{single-link} \emph{LTE} throughput of 86~Mbps; the \emph{LAA} entrant achieves a slightly lower
throughput than \emph{LTE}, of 78~Mbps, due to the LBT MAC overhead term $S$; the \emph{\mbox{LTE-U} fixed 50\% duty
cycle} achieves exactly half the \emph{LTE} throughput; and the \emph{\mbox{Wi-Fi}} entrant achieves the lowest throughput
of 37~Mbps due to its lower PHY layer efficiency than \emph{LTE} (and slightly higher LBT MAC overhead
than \emph{LAA}). We note that the median throughput results for the \emph{indoor/indoor} and \emph{indoor/outdoor}
scenarios, for all considered incumbent network densities, are identical to those in Fig.~\ref{fig_4}\subref{fig_4a}.

The coexistence mechanism implemented by the \mbox{LTE-in-unlicensed} variant only starts to play a role
in the achieved entrant throughput once the likelihood of \mbox{co-channel} operation by entrant APs within
CS range increases significantly,\footnote{We note that this differs from the incumbent AP throughput results in Fig.~\ref{fig_3}, because the entrant AP using \emph{sense} channel selection will avoid operating in a channel occupied by an incumbent AP if possible, but is more likely to randomly select the same channel as another entrant AP (once the number of entrant APs is high enough).} 
as demonstrated by the 10\textsuperscript{th} percentile entrant throughput in
Figs.~\ref{fig_4}\subref{fig_4b} and~\ref{fig_4}\subref{fig_4c}. Specifically, Fig.~\ref{fig_4}\subref{fig_4b} shows that the ordering of the LTE variant curves changes
once the entrant network density reaches a critical value (of around \mbox{5--7} APs) where the likelihood of
\mbox{co-channel} operation becomes significant, and mechanisms to avoid interference thus become useful.
The results in Fig.~\ref{fig_4}\subref{fig_4c} show that this \emph{critical network density} is only around 1--2 entrant APs with a
higher transmission power $P_{tx}$, due to a higher corresponding CS range, and thus higher likelihood of
\mbox{co-channel} interference for neighbouring APs. We note that we observed a similar trend of \emph{two distinct
regimes for entrant variant ranking} also for other scenarios, but with different critical
network densities.

In this second regime beyond the critical network density, where interference starts to occur, the
\emph{\mbox{LTE-U} ideal} variant achieves the highest entrant AP throughput, consistent with its superior PHY
layer and maximum MAC layer coordination. The \emph{LAA} entrant achieves only a slightly lower
throughput than \emph{\mbox{LTE-U} ideal} due to the MAC layer overhead of LBT. By contrast, the other LBT
entrant \emph{\mbox{Wi-Fi}} achieves the lowest throughput of all entrant variants, simply due to a lower PHY layer
spectral efficiency than \emph{LAA}.\footnote{We note \emph{\mbox{Wi-Fi}} LBT also implements a more conservative CS threshold than \emph{LAA}. However, we have observed in our other simulation results~\cite{voicu2016} that \emph{\mbox{Wi-Fi}}'s lower CS threshold is somewhat beneficial in the \emph{outdoor/outdoor} scenario; nonetheless, it is important to choose a \mbox{well-tuned} CS threshold to maximize network throughput of LBT technologies in shared spectrum.} 
Finally, the \emph{LTE} and \emph{\mbox{LTE-U} duty cycle} variants exhibit a decreasing
entrant throughput with increasing (\mbox{SINR-reducing}) interference in dense networks, consistent with
their implementing either no or ineffective interference coordination mechanisms at the MAC layer.
It is therefore only in this \emph{regime of likely interference}---i.e. situations when \mbox{co-channel} APs are
situated within CS range of each other---that distributed interference coordination, as facilitated by
the LBT MAC mechanism of the \emph{\mbox{Wi-Fi}} incumbent, becomes necessary to prevent a ``tragedy of the
commons'' in an unlicensed spectrum band.

\subsection{Worst-Case Impact of Entrant on Wi-Fi Incumbent}
\label{worst-case impact}

Finally, let us consider the impact of different \mbox{LTE-in-unlicensed} entrant variants on the incumbent
throughput in situations where the incumbent APs are also affected by the \emph{regime of likely
interference} observed for the entrants in Fig.~\ref{fig_4} after a critical network density. We emphasize that
although our results in Fig.~\ref{fig_3} demonstrate that such situations are expected to occur infrequently,
they are nonetheless important, as even such rare cases of coexistence problems have the potential to
motivate calls for regulatory intervention (as discussed in Section~\ref{ent_tech}). To this end, Fig.~\ref{fig_5} shows
the \emph{\mbox{Wi-Fi}} incumbent AP throughput in the \emph{indoor/indoor} scenario for the \mbox{worst-case} coexistence
situation where the incumbent and entrant APs all operate on a single channel (i.e. \emph{forced \mbox{co-channel}}).

We observe from Fig.~\ref{fig_5} that the \emph{LTE} entrant, which implements no coexistence mechanism
whatsoever, results in a severe deterioration of incumbent performance, reducing the median
incumbent AP throughput to zero when there are more than 6 \mbox{co-channel} entrant APs and for all
entrant densities in the 10\textsuperscript{th} percentile case. Thus Fig.~\ref{fig_5} makes it clear that some kind of coexistence
mechanism is necessary to avoid disrupting the incumbent. However, Fig.~\ref{fig_5} also shows that the
difference in impact on the incumbent's performance among the other \mbox{LTE-in-unlicensed} variants
(i.e. \emph{LAA} and \emph{\mbox{LTE-U}}) is typically under 1~Mbps, and at most 3~Mbps. This result has important
policy implications, as it suggests that, contrary to arguments and claims made in many recent heated
industry and regulatory debates~\cite{lteuforum2015, qualcomm2015, wifialliance2015, fcc2015}, in practice it does not make a difference to the incumbent
what kind of coexistence mechanism is added to \emph{LTE} operating in unlicensed bands, as long as one is
in place. In other words, the debate over whether LAA should be mandatory or \mbox{LTE-U} may be
allowed~\cite{fcc2015}, is immaterial to the policy goal of ensuring coexistence with the \mbox{Wi-Fi} incumbent; the
choice of LAA or \mbox{LTE-U} should be made by the \mbox{LTE-in-unlicensed} operator, based on the resulting
performance for the entrant, and is of no concern for the spectrum regulator.

Importantly, Fig.~\ref{fig_5} answers our original motivating research question: LTE is neither friend nor foe
to \mbox{Wi-Fi} in the unlicensed bands \emph{in general}. Fig.~\ref{fig_5}\subref{fig_5a} shows that in some cases the \mbox{LTE-in-unlicensed}
variants are indeed better neighbours to the \emph{\mbox{Wi-Fi}} incumbent than the baseline \emph{\mbox{Wi-Fi}}
entrant itself. This is consistent with existing claims by e.g. the LTE-U forum~\cite{lteuforum2015, qualcomm2015}. The opposite
holds in the cases of Figs.~\ref{fig_5}\subref{fig_5b} and~\ref{fig_5}\subref{fig_5d}. This result stems simply from the fact that the \emph{\mbox{Wi-Fi}}
incumbent applies a lower CS threshold when deferring to \emph{\mbox{Wi-Fi}} than when deferring to another
technology (e.g. the \mbox{LTE-in-unlicensed} entrants), of -82~dBm and -62~dBm, respectively. The lower
CS threshold is poorly tuned for networks with low effective density, and well tuned for dense
networks. Accordingly, in Fig.~\ref{fig_5}\subref{fig_5a} the \emph{\mbox{Wi-Fi}} incumbent is unnecessarily ``polite'' to the \emph{\mbox{Wi-Fi}}
entrant, to the detriment of its throughput, and achieves a higher median throughput when it defers
less to equivalently far away \mbox{LTE-in-unlicensed} entrants. On the other hand, a higher level of
politeness starts to pay off in the cases of higher effective node density in Figs.~\ref{fig_5}\subref{fig_5b}-\subref{fig_5d}: the
likelihood of a large number of nearby APs causing harmful interference to the incumbent increases
in the order of Fig.~\ref{fig_5}\subref{fig_5c} to Fig.~\ref{fig_5}\subref{fig_5b} to Fig.~\ref{fig_5}\subref{fig_5d}, matched by the trend of increasing superiority of
\emph{\mbox{Wi-Fi}}'s coexistence with itself compared to with \mbox{LTE-in-unlicensed}. In the most extreme case of the
10\textsuperscript{th} percentile throughput of 10 incumbent APs coexisting with 10 entrant APs shown in Fig.~\ref{fig_5}\subref{fig_5d},
all \mbox{LTE-in-unlicensed} variants completely prevent the \emph{\mbox{Wi-Fi}} incumbent from transmitting, whereas
the \mbox{all-\emph{Wi-Fi}} network still achieves a throughput of around 5~Mbps.

Finally, we note that for all \emph{outdoor/outdoor} scenarios we observed the same qualitative results as in
Fig.~\ref{fig_5}, whereas for the \emph{indoor/outdoor} scenarios we observed the same results as in Fig.~\ref{fig_3}, since the
entrant and incumbent networks remain independent even in the \emph{forced \mbox{co-channel}} case due to the
isolation afforded by high external building material losses~\cite{voicu2015}.

\clearpage

\begin{figure}[t!]
\centering
\subfloat{\includegraphics[scale=0.62]{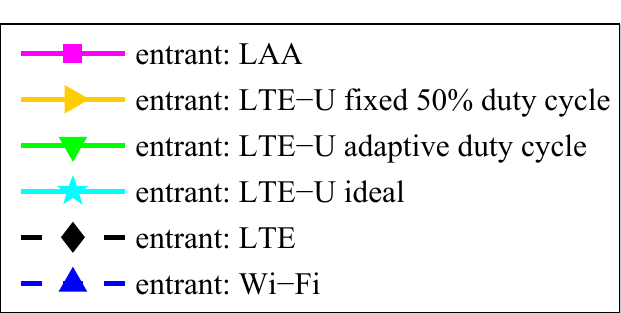}}
\\  
\captionsetup[subfigure]{width=0.5\linewidth, font=scriptsize}
\setcounter{subfigure}{0}         
\subfloat[\textbf{Median} incumbent throughput for \textbf{1 incumbent AP}.]
          {\includegraphics[scale=0.57]{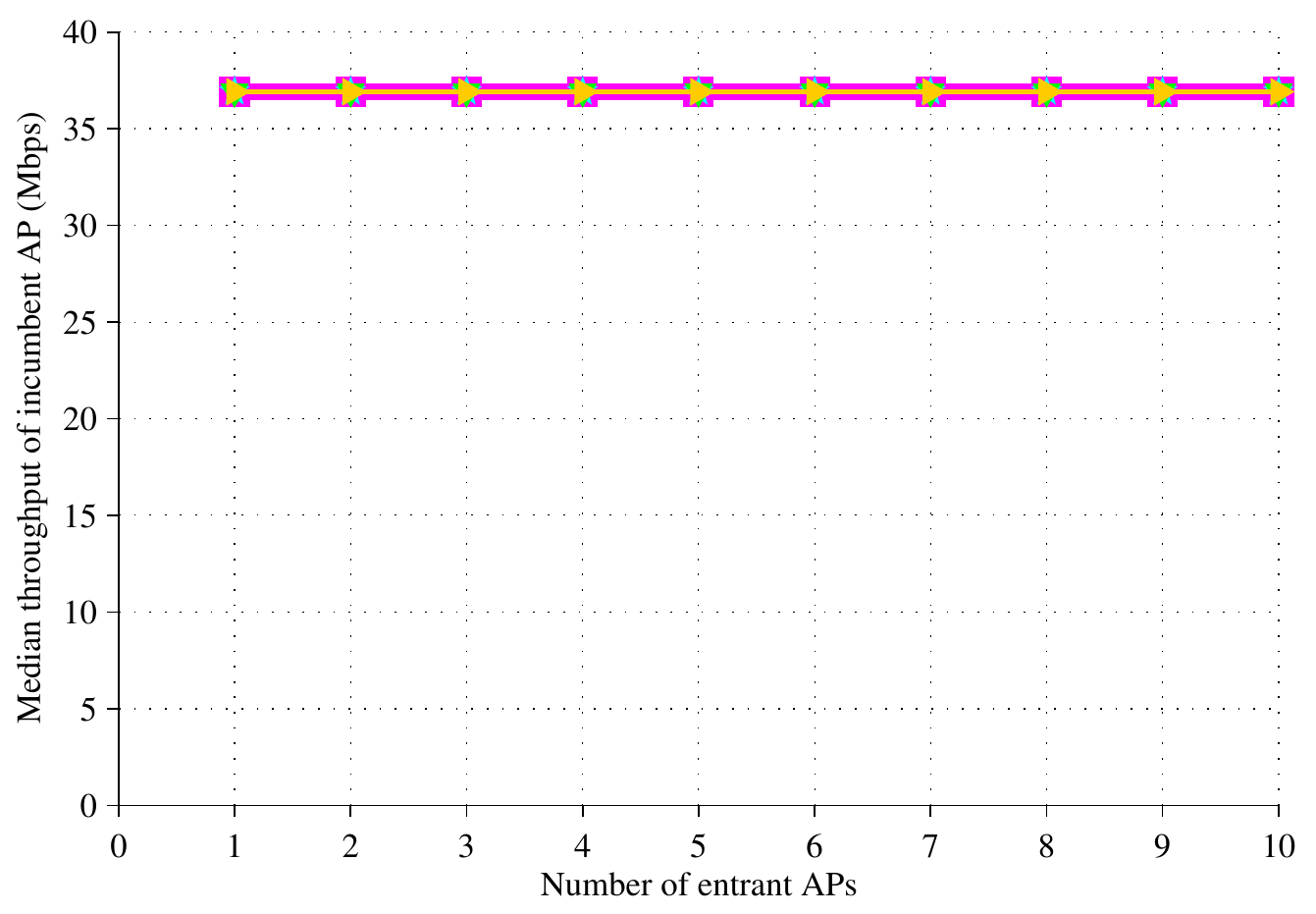} \label{fig_3a}}
~
\subfloat[\textbf{10\textsuperscript{th} percentile} incumbent throughput for \textbf{1 incumbent AP}.]   
	  	   {\includegraphics[scale=0.57]{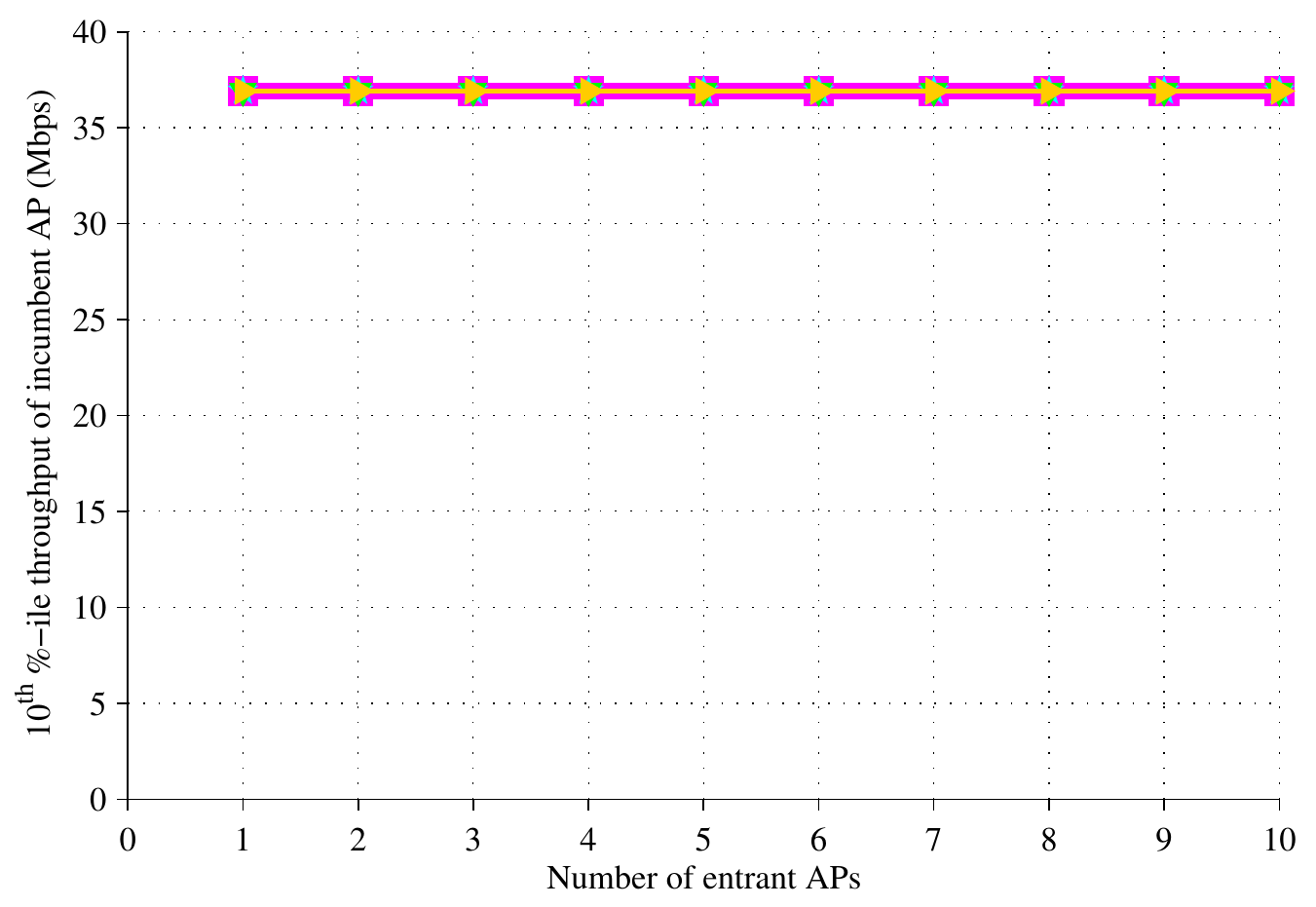} \label{fig_3b}}
\\
\subfloat[\textbf{Median} incumbent throughput for \textbf{10 incumbent APs}.]
          {\includegraphics[scale=0.57]{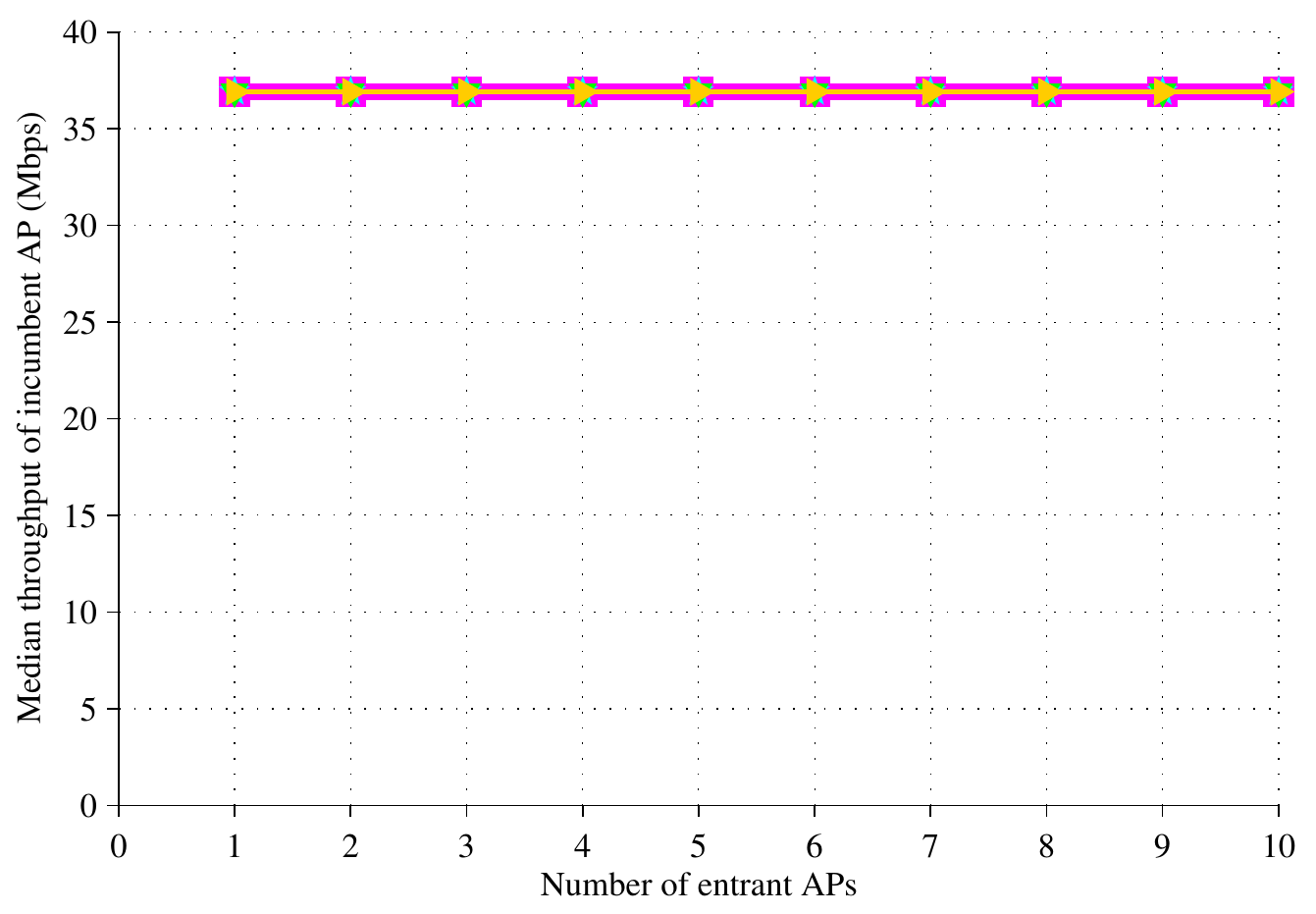} \label{fig_3c}}
~
\subfloat[\textbf{10\textsuperscript{th} percentile} incumbent throughput for \textbf{10 incumbent APs}.]      	  	   		       		 
          {\includegraphics[scale=0.57]{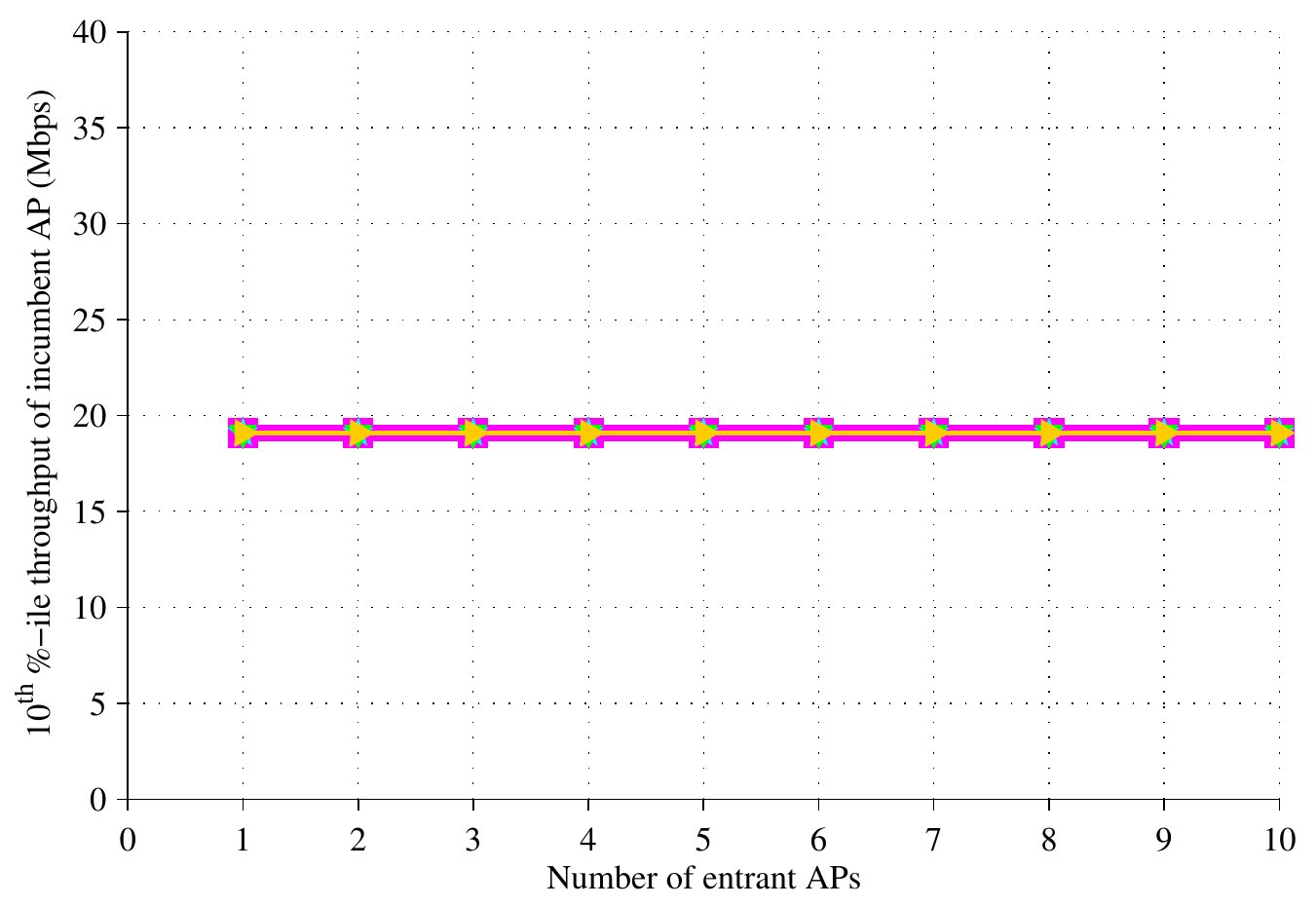} \label{fig_3d}}
\caption{Median and 10\textsuperscript{th} percentile throughput of incumbent \emph{\mbox{Wi-Fi}} AP when coexisting with different
entrant technologies, for the \emph{indoor/indoor} scenario with \emph{sense} entrant channel selection of 1 of 19
indoor channels. (The trend of constant incumbent AP throughput, regardless of entrant number or
variant, shown here, was also observed in our simulation results for the \emph{indoor/outdoor} and
\emph{outdoor/outdoor} scenarios.)}
\label{fig_3}
\end{figure} 

\clearpage

\begin{figure}[t!]
\centering
\subfloat{\includegraphics[scale=0.62]{legend_10}}
\\  
\captionsetup[subfigure]{width=\linewidth, font=scriptsize}
\setcounter{subfigure}{0}         
\subfloat[\textbf{Median} entrant throughput, with all APs transmitting at $\boldsymbol{P_{tx}}$\textbf{=23~dBm}.]
          {\includegraphics[scale=0.57]{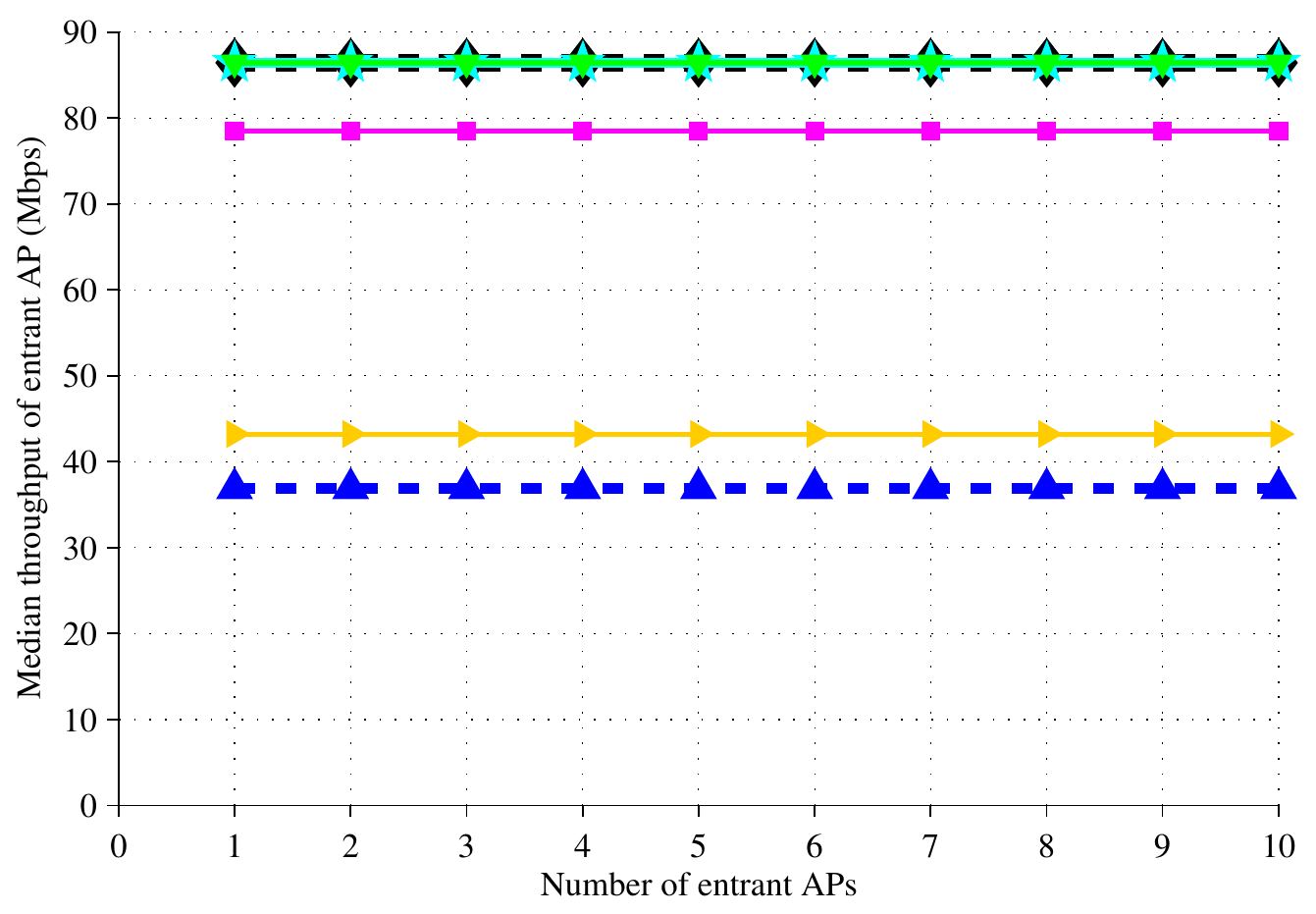} \label{fig_4a}}
\\
\subfloat[\textbf{10\textsuperscript{th} percentile} entrant throughput, with all APs transmitting at $\boldsymbol{P_{tx}}$\textbf{=23~dBm}.]   
	  	   {\includegraphics[scale=0.57]{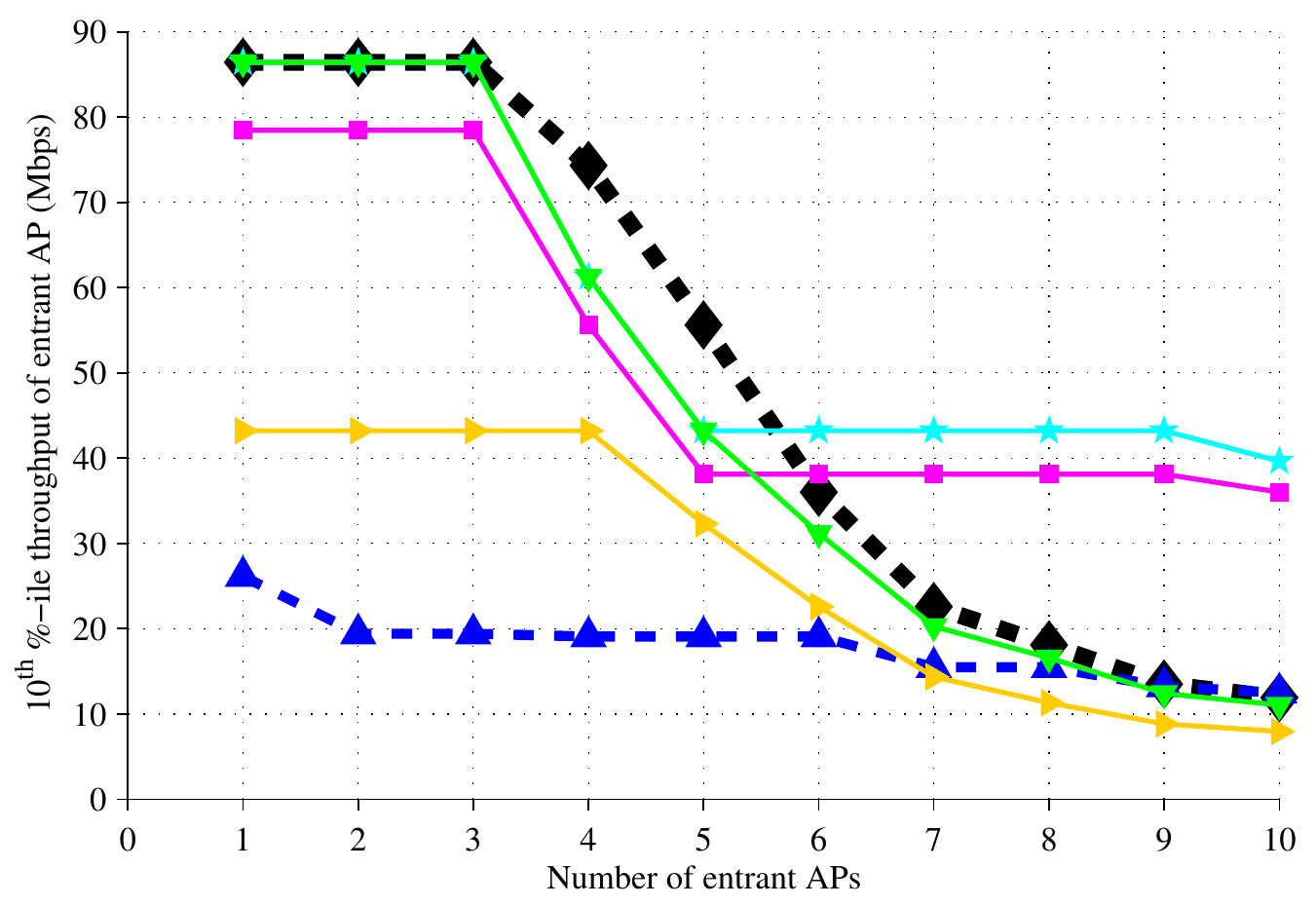} \label{fig_4b}}
\\
\subfloat[\textbf{10\textsuperscript{th} percentile} entrant throughput, with all APs transmitting at $\boldsymbol{P_{tx}}$\textbf{=30~dBm}.]
          {\includegraphics[scale=0.57]{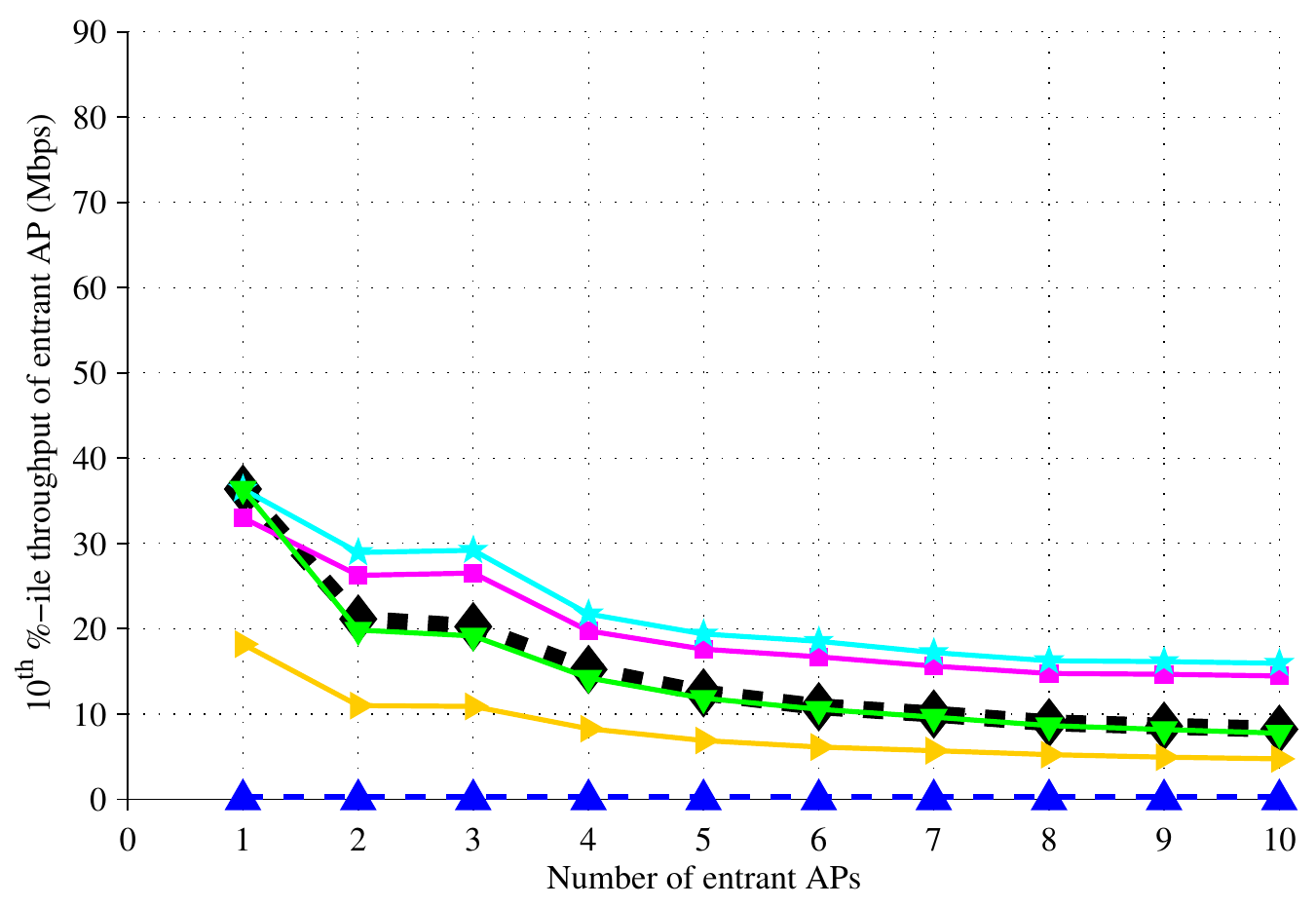} \label{fig_4c}}
\caption{Median and 10\textsuperscript{th} percentile throughput of entrant AP coexisting with 1 incumbent \emph{\mbox{Wi-Fi}} AP, for
different entrant technologies, for the \emph{outdoor/outdoor} scenario with \emph{sense} selection of one of 11 outdoor channels, and different AP transmit powers.}
\label{fig_4}
\end{figure}

\clearpage

\begin{figure}[t!]
\centering
\subfloat{\includegraphics[scale=0.62]{legend_10}}
\\  
\captionsetup[subfigure]{width=0.5\linewidth, font=scriptsize}
\setcounter{subfigure}{0}         
\subfloat[\textbf{Median} incumbent throughput for \textbf{1 incumbent AP}.]
          {\includegraphics[scale=0.57]{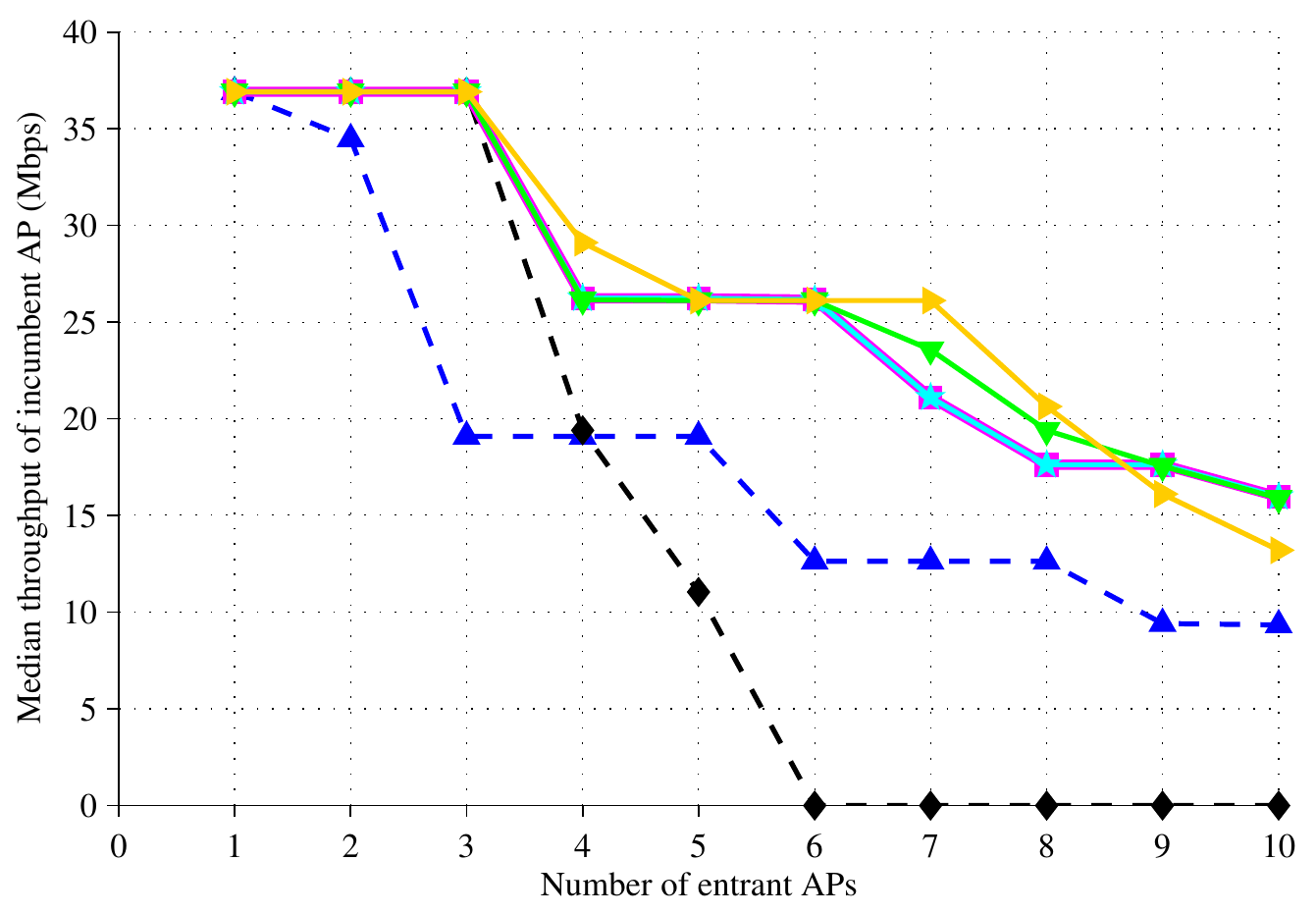} \label{fig_5a}}
~
\subfloat[\textbf{10\textsuperscript{th} percentile} incumbent throughput for \textbf{1 incumbent AP}.]   
	  	   {\includegraphics[scale=0.57]{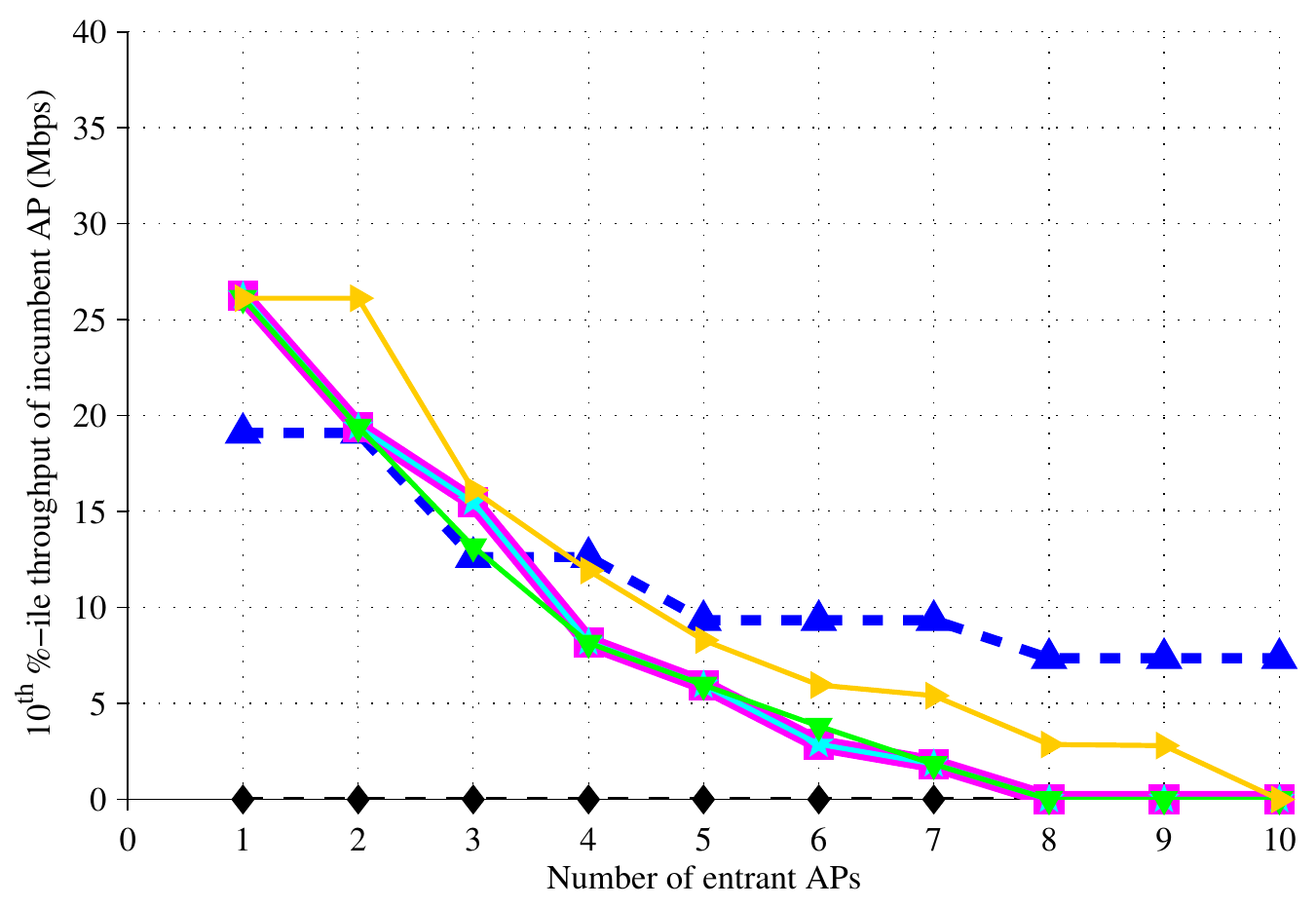} \label{fig_5b}}
\\
\subfloat[\textbf{Median} incumbent throughput for \textbf{10 incumbent APs}.]
          {\includegraphics[scale=0.57]{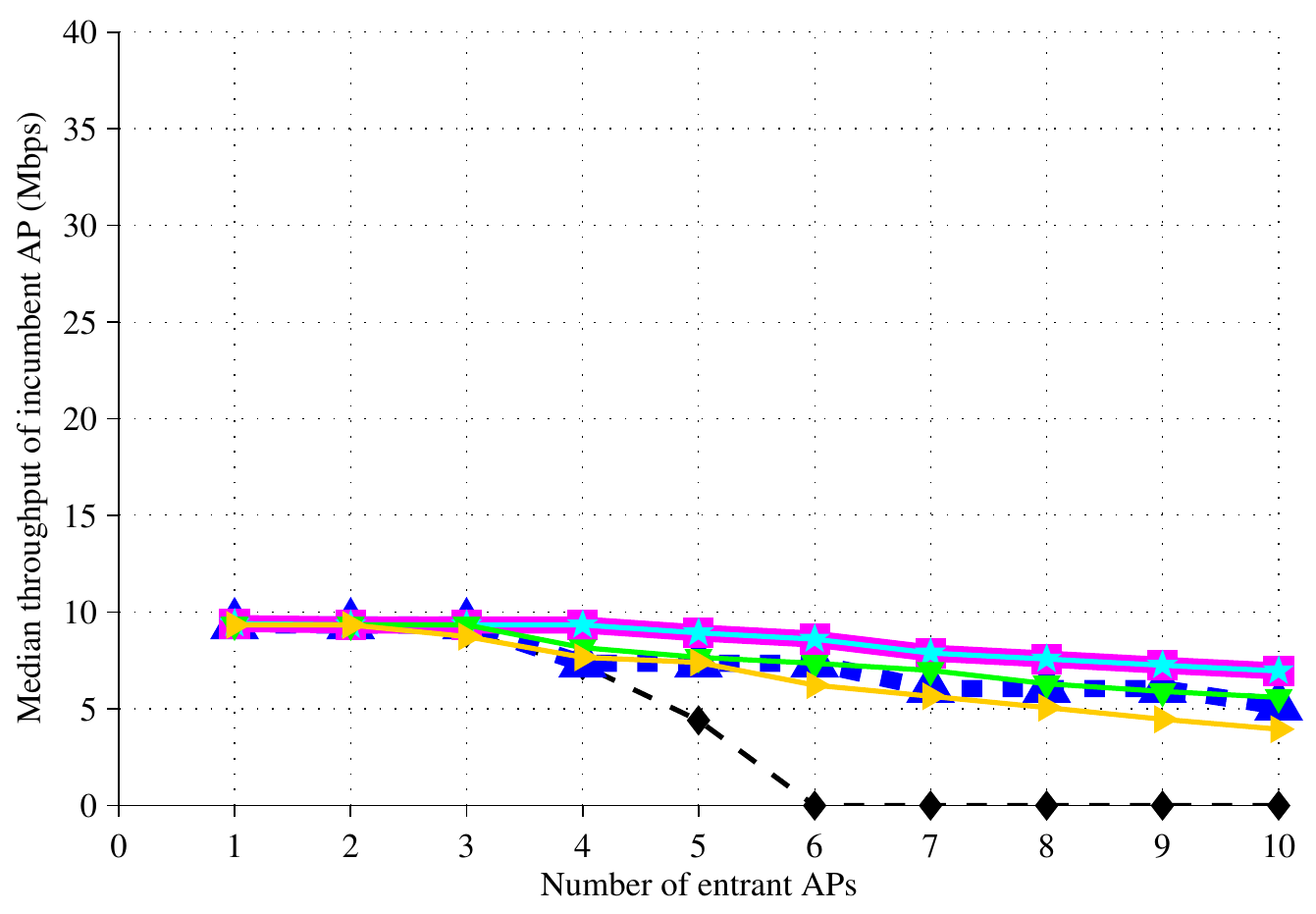} \label{fig_5c}}
~
\subfloat[\textbf{10\textsuperscript{th} percentile} incumbent throughput for \textbf{10 incumbent APs}.]      	  	   		       		 
          {\includegraphics[scale=0.57]{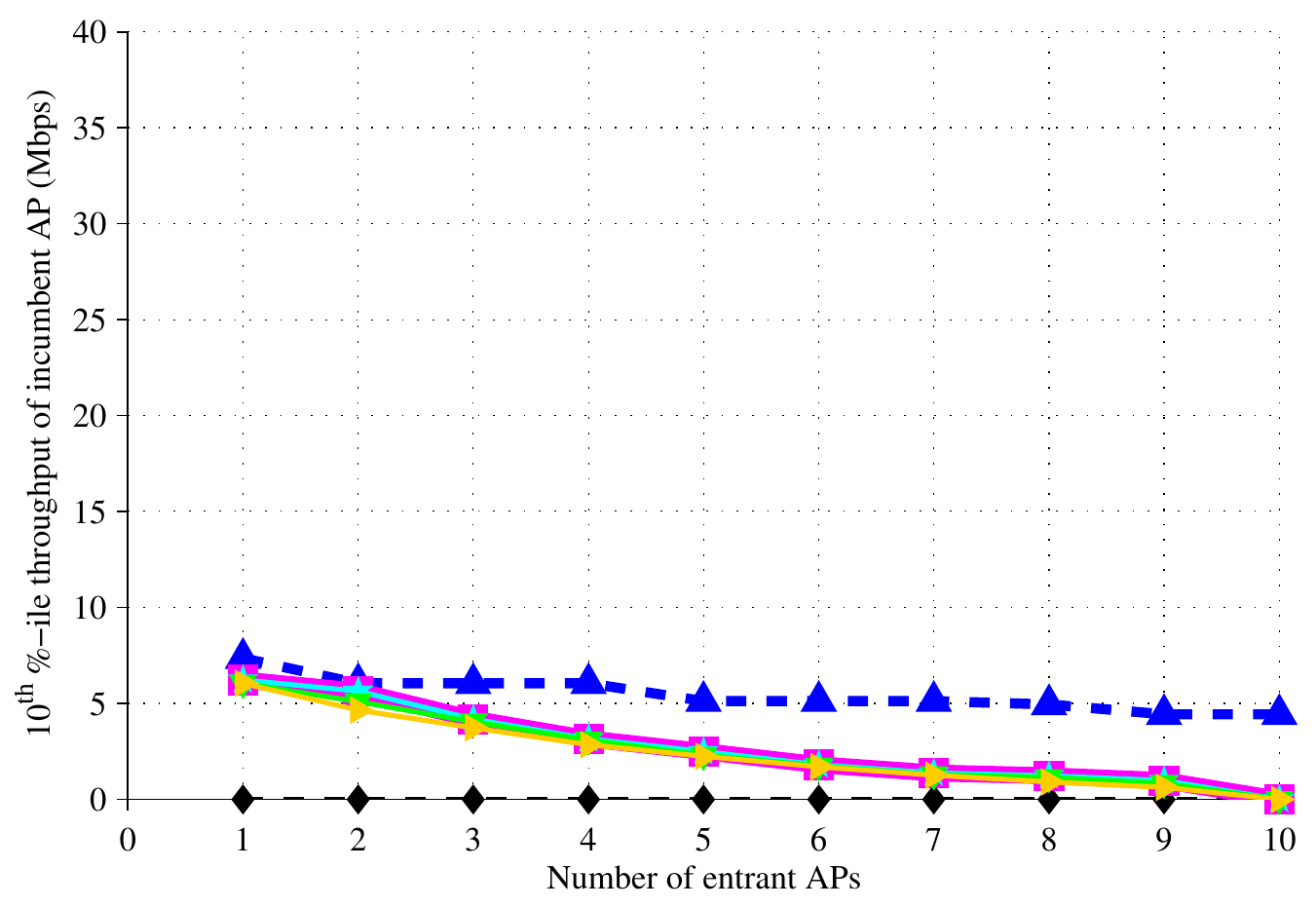} \label{fig_5d}}
\caption{Median and 10\textsuperscript{th} percentile throughput of incumbent AP when coexisting with different entrant
technologies, for the \emph{indoor/indoor} scenario with \emph{forced co-channel} operation of all APs.}
\label{fig_5}
\end{figure} 

\clearpage

\section{Conclusions}

We presented a coexistence study of \mbox{Wi-Fi} and LTE in the 5~GHz unlicensed band, surveying a large
parameter space of coexistence mechanisms and a range of network scenarios. We thereby sought to
identify whether situations warranting regulatory intervention to protect the incumbent \mbox{Wi-Fi}
technology from the new LTE entrant are likely to arise. Our results show that, for typical network
densities, harmonious coexistence between \mbox{Wi-Fi} and LTE is ensured by the large number of 5~GHz
channels which mean that \mbox{co-channel} operation is easily avoided. For the \mbox{worst-case} scenario of
\emph{forced \mbox{co-channel}} operation, LTE is sometimes a better neighbour to \mbox{Wi-Fi} than a \mbox{Wi-Fi} entrant---
when effective node density is low---but sometimes worse---when the density, and the potential for
interference, is high. We also showed that it does not make a difference to the \mbox{Wi-Fi} incumbent
which LTE coexistence mechanism is implemented (i.e. \mbox{LTE-U} or LAA), as long as one is in place.
Therefore, we conclude that LTE is neither friend nor foe to \mbox{Wi-Fi} in the unlicensed bands \emph{in
general}, contrary to the claims of both lobbying camps. We argue that the systematic engineering
analysis demonstrated by our \mbox{Wi-Fi}/LTE case study is a \mbox{best-practice} approach for supporting
\mbox{evidence-based} rulemaking by the regulator.


%





\ifCLASSOPTIONcaptionsoff
  \newpage
\fi

\end{document}